\documentclass[10pt,journal,compsoc]{IEEEtran}

\ifCLASSOPTIONcompsoc
  \usepackage[nocompress]{cite}
\else
  \usepackage{cite}
\fi

\usepackage{amsmath,amssymb,amsfonts}
\usepackage{algorithmic}
\usepackage{graphicx}
\usepackage{textcomp}
\usepackage[table]{xcolor}
\usepackage{tabularx}
\usepackage{booktabs}
\usepackage[hidelinks]{hyperref}
\newcommand{\C}[1]{{\footnotesize \textsf{#1}}}

\definecolor{VeryLightGray}{rgb}{0.92,0.92,0.92}
\usepackage{colortbl}
\newcommand{\graybg}{\cellcolor{VeryLightGray}}

\usepackage{nicefrac}

\usepackage{etoolbox}
\AtBeginEnvironment{quote}{\vspace{0.5\baselineskip}}
\AtEndEnvironment{quote}{\vspace{0.5\baselineskip}}

\begin{document}

\title{Contextual Documentation Referencing\\on Stack~Overflow}

\author{
Sebastian~Baltes, %~\IEEEmembership{Member,~IEEE,}
Christoph~Treude, %~\IEEEmembership{Member,~IEEE,}
Martin~P.~Robillard%,~\IEEEmembership{Member,~IEEE}%
\IEEEcompsocitemizethanks{%
\IEEEcompsocthanksitem Sebastian Baltes is with The University of Adelaide, Australia.\protect\\
E-mail: sebastian.baltes@adelaide.edu.au%
\IEEEcompsocthanksitem Christoph Treude is with The University of Adelaide, Australia.\protect\\
E-mail: christoph.treude@adelaide.edu.au%
\IEEEcompsocthanksitem Martin P. Robillard is with McGill University, Montréal, Canada.\protect\\
E-mail: martin@cs.mcgill.ca%
}%
\thanks{Manuscript received March 11, 2019.}}

\markboth{IEEE~TRANSACTIONS~ON~SOFTWARE~ENGINEERING,~Vol.~X, No.~X, March~2019}%
{Baltes \MakeLowercase{\textit{et al.}}: Contextual Documentation Referencing on Stack Overflow}

\IEEEtitleabstractindextext{%
% Include: Context-Objective-Method-Results-Conclusion
\begin{abstract}
%\todo{R2: Please, revise it to make it easier to understand. In particular, try to provide a little more details about the findings (i.e., avoid writing "a range of different purposes" without mentioning a few examples at least). The last sentence is quite long and convoluted.}
Software engineering is knowledge-intensive and requires software developers to continually search for knowledge, often on community question answering platforms such as Stack Overflow.
Such information sharing platforms do not exist in isolation, and part of the evidence that they exist in a broader software documentation ecosystem is the common presence of hyperlinks to other documentation resources found in forum posts.
With the goal of helping to improve the information diffusion between Stack Overflow and other documentation resources, we conducted a study to answer the question of how and why documentation is referenced in Stack Overflow threads.
We sampled and classified 759 links from two different domains, regular expressions and Android development, to qualitatively and quantitatively analyze the links' context and purpose, including attribution, awareness, and recommendations.
We found that links on Stack Overflow serve a wide range of distinct purposes, ranging from citation links attributing content copied into Stack Overflow, over links clarifying concepts using Wikipedia pages, to recommendations of software components and resources for background reading.
This purpose spectrum has major corollaries, including our observation that links to documentation resources are a reflection of the information needs typical to a technology domain.
We contribute a framework and method to analyze the context and purpose of Stack Overflow links, a public dataset of annotated links, and a description of five major observations about linking practices on Stack Overflow.
Those observations include the above-mentioned purpose spectrum, its interplay with documentation resources and applications domains, and the fact that links on Stack Overflow often lack context in form of accompanying quotes or summaries.
We further point to potential tool support to enhance the information diffusion between Stack Overflow and other documentation resources.
%, with detailed links to evidence, implications, and a conceptual framework to capture the relations between the five observations.
\end{abstract}

}

\maketitle

\IEEEdisplaynontitleabstractindextext

\IEEEpeerreviewmaketitle

\IEEEraisesectionheading{\section{Introduction}
\label{s:intro}}

\IEEEPARstart{T}{he} knowledge-intensive nature of current-day software engineering means that software developers are continually in search of knowledge. A popular model for knowledge sharing on the Internet is the community question answering site, with Stack Overflow~\cite{SPO2008a} serving as the de facto forum for most programmers~\cite{MMM2011a}. On Stack Overflow, registered users can post questions, answer posted questions, and comment on questions and answers by other users, which can then be viewed by anyone.
% retrieved 2019-12-23 using https://data.stackexchange.com/stackoverflow/query/new
As of December 2019, Stack Overflow archives 19M questions, 28M answers, and 72M comments. At this scale, Stack Overflow constitutes a major information broker between posters, contributors, and non-contributing readers (so-called ``lurkers'').

% Stack Overflow lives in a documentation ecosystem
Stack Overflow, however, does not exist in isolation---the site is only one of many sources of programmer knowledge in a software documentation ecosystem. Past research has extensively characterized the strengths and weaknesses of Stack Overflow (e.g., good at ``how-to'' documentation~\cite{TreudeBarzilayOthers2011}, bad at completeness~\cite{Parnin2012}) compared to other sources, such as API documentation (e.g., good at structure~\cite{Maalej2013}, bad at scenarios~\cite{Robillard2011}). Meng et al.'s observational study corroborates that developers seek a diversity of documentation content when solving programming tasks~\cite{Meng2018-1}.

With these complementary strengths and weaknesses, it is only natural that links exist from one source to another. In fact, previous studies found that link sharing is a significant phenomenon on Stack Overflow that make the site part of a larger interconnected network of online resources used and referenced by developers~\cite{GCS2013a, VJH2018a}. Given the crucial role that on-line resources play in developers' quest for technical knowledge, it is important to know how information is diffused between resources types so we can facilitate this quest (see Section~\ref{s:examples}).

% Move this to the background section

We conducted a multi-case study to answer the question of \textit{how and why documentation is referenced in Stack Overflow threads}. We sampled 759 links from two different domains (Java regular expressions and Android development), classified and qualitatively analyzed them, and then used the resulting data to derive association rules and build logistic regression models to identify properties of Stack Overflow questions that attract links to documentation resources.

Our main findings include that links on Stack Overflow serve widely diverse purposes that range from simple pointers to API documentation over links to concept descriptions on Wikipedia to suggestions of software components and background readings. This \emph{purpose spectrum} (see Section~\ref{s:results}) allows us to modulate Stack Overflow's requirement to add context for links~\cite{SO2019a}. We also find that links to documentation resources are a reflection of the information needs typical to a technology domain, with significant differences between the two domains in our multi-case study.

Our main contributions are: (1) a framework and method to analyze the \emph{context} and \emph{purpose} of documentation links on Stack Overflow, (2) a public dataset with 759 annotated links that other researchers can use, and (3) a description of five major observations about linking practices on Stack Overflow, with detailed links to evidence, implications, and a conceptual framework to capture the relations between the five observations.

The remainder of this paper is structured as follows: We provide additional background and motivation in Section~\ref{s:examples} and outline our study design in Section~\ref{s:design}. Section~\ref{s:sampling} describes our method for link sampling and classification, Sections~\ref{s:qualitative} and \ref{s:quantitative} describe our qualitative and quantitative analyses, respectively. Section~\ref{s:results} presents the major findings derived from these analyses, Section~\ref{s:threats} describes threats to validity. We conclude the paper in Section~\ref{s:conclusion}.

\section{Background and Motivation}
\label{s:examples}

This work is a systematic investigation of current \textbf{information diffusion (link sharing) practices on Stack Overflow}, with the goal of informing the development of \textbf{advanced technology to facilitate this diffusion}. This research takes place in the context of \textbf{previous studies on information diffusion in on-line developer communities}.

\subsection*{Information Diffusion on Stack Overflow}

Stack Overflow explicitly encourages the inclusion of links to external resources in answers, but requests that users add context so that ``fellow users will have some idea what it is and why it's there.''~\cite{SO2019a}. This advice is overly general. Not all link targets need to be quoted, and in some cases, the context for a link is obvious. However, deciding when and how to include links to other documentation sources in Stack Overflow posts requires differentiating common linking practices and understanding their unique characteristics. The following examples illustrate the richness and diversity of linking practices on Stack Overflow.

When considering the potential value of links on Stack Overflow, the best case scenario is the recommendation of specific information relevant to the thread (links are in bold):

\begin{quote}
\sf
\footnotesize
...have a look at \textbf{Greedy, Reluctant, and Possessive Quantifiers} section of the Java RegEx tutorial...~\cite{SO1}
% ROW 247: https://stackoverflow.com/q/21761788#comment32918966_21761788
\end{quote}

In this case, a contributor provided a comment to point the original poster to a section of a tutorial introducing the concept of regular expression quantifiers and explaining how to use them. These ``ideal'' links provide clear value added to the thread, and form a type of information that can even be automatically mined to improve information discovery~\cite{LXS2018}.

However, the reality of linking practices goes broadly beyond this expected scenario. For example, links to obvious documentation resources can be introduced defensively by the original poster themselves, to avoid having a question \emph{downvoted}~\cite{Sle2015}:

\begin{quote}
\sf
\footnotesize
%I am facing a weird problem with Java Regex... Read through \textbf{Oracle - RegexBounds} and \textbf{RegularExpressions - WordBoundaries}...\cite{SO2}
% ROW 53: https://stackoverflow.com/q/27788811
I've already tried this solution (\textbf{http://developer.android .com/training/articles/security-ssl.html}) but I still have the same error:...\cite{SO29}
\end{quote}

Other links bind a reference to library classes to its documentation. This can be useful to help make code fragments more self-explanatory~\cite{TreudeRobillard2017}, but we observed that such links are also provided for well-known, pervasive classes:

\begin{quote}
\sf
\footnotesize
When you want to return more than one result, you need to return an array (String[]) or a Collection like an \textbf{ArrayList}, for example.\cite{SO3}
% ROW 40: https://stackoverflow.com/a/24890987
\end{quote}

From the point of view of links as mechanisms to increase the flow of valuable software development knowledge, degenerate practices include providing links to comic strips (such as xkcd) and similar sites:

\begin{quote}
\sf
\footnotesize
...reminds me of this \textbf{xkcd}\cite{SO4}
% https://stackoverflow.com/questions/30910685/java-regular-expression-to-discover-regular-expression
\end{quote}

% \noindent
% Links to the profile of other contributors in the same thread (a courtesy that turns the link into a clear false positive in terms of source of development information):
%
% \begin{quote}
%
% \end{quote}

% And, arguably one of the most resented pieces of information on the site, the inclusion of the link to a famous placating blog post~\cite{Gem2008a}:
%
% \begin{quote}
% \sf
% \footnotesize
% I like to refer [you] to \textbf{whathaveyoutried.com}...\cite{SO5}
% % https://stackoverflow.com/questions/15565774/does-anyone-know-how-extract-my-date-string-change-the-format
% \end{quote}

As these examples show, linking practices on Stack Overflow are diverse and the intrinsic value of a link as a carrier of relevant technical information is not uniform. The first example link, to a specific section of a tutorial, has an obvious purpose and value. The link to a comic strip is clearly noise. Between these extremes lies a gray zone where links play different roles in different contexts.

\subsection*{Enhancing Information Diffusion}

As illustrated above, links in on-line developer forums can fulfill the important mission of complementing documentation with explanations of concepts or descriptions of code elements. However, a manual linking process is prone to omissions. A number of techniques have been proposed to automatically enhance on-line resources through linking and recommendation.

The idea of automatically enhancing information diffusion is clearly captured by Gao et al.'s proposal to automatically add links to recognized entities in Stack Overflow posts from a database of popular URLs, and taking into account the context in which the entity appears~\cite{GXM2017a}. A different take on the problem is offered by Li et al.~\cite{LXS2018}, who built a collaborative filtering recommender system to recommend other learning resources, based on co-occurrences of links in Stack Overflow posts. In approaches that are based on existing link data, the automatic linking system relies on the assumption that the underlying linking practices are sound. Our study sheds light on the linking practices that are used as foundations for collaborative filtering. Xu et al.'s deep-learning-based approach for predicting semantically linkable knowledge in developer forums~\cite{XYX2016a} avoids the issue of relying on existing links. This is an important advancement for improving information diffusion in knowledge networks. However linking that is based on the semantics of the text may not necessarily take into account the \textit{purpose} for linking a knowledge unit (e.g., the comic strip mentioned above). Our study focuses specifically on eliciting the purpose of links so that it is possible to account for it when enabling information diffusion though automated approaches.

% I don't get the point of this part
% To describe topics of Stack Overflow questions and answers, different methods such as manual analysis~\cite{TreudeBarzilayOthers2011} and Latent Dirichlet Allocation~\cite{WangLoDavidOthers2013, AllamanisSutton2015} have been used.

Content is one concern for documentation ecosystems, but quality is another important one. Previous work has attempted to automatically identify high-quality posts using features based on the number of edits on a question~\cite{YangHauffOthers2014}, author popularity~\cite{PonzanelliMocciOthers2014}, and code readability~\cite{DuijnKuceraOthers2015}. In their conceptual framework of success factors for Stack Overflow questions, Calefato et al.~\cite{CalefatoLanubileOthers2018} considered the presence of links as one aspect of a question's presentation quality. However, they did not find a significant effect of the fact that a question contained a link on the success of that question, that is whether it attracted an accepted answer. A direction of future work is to consider not only the presence of a links, but also their purpose and targets, as enabled by our study.

\subsection*{Studies of Information Diffusion}

There have been different studies investigating individual aspects of link usage on Stack Overflow. Gomez et al.~\cite{GCS2013a} conducted a preliminary study of the links found on Stack Overflow. Their study focused on the different types of links in posts (not comments) and it did not factor in a distinction based on the domain. In this article, we investigate two specific domains, which allows us to understand the data in a specific context. Moreover, we integrate an analysis of the purpose of the information sharing that goes beyond a basic description of its nature.

Vincent et al.~\cite{VJH2018a} analyzed the usage of Wikipedia by Stack Overflow authors. They found that 1.28\% of all Stack Overflow posts contain links to Wikipedia. Using version 2018-07-31 of the \emph{SOTorrent}~dataset~\cite{BDTD2018}, we identified 1.94\% of all threads, but only 0.85\% of all posts, to contain links to Wikipedia.
Also considering links in comments, which Vincent et al. did not, the ratio of threads with links to Wikipedia increases to 2.55\%.
%\todo{SB: I'm pretty sure that Vincent et al. mixed up x \% of all posts with x \% of all threads. Is this paragraph too offensive towards Vincent et al.?}

Ye et al. contributed a study of link sharing on Stack Overflow that focuses on the sharing of links to other Stack Overflow posts\cite{YXK2017a}. In contrast, our study covers links to all external resources, and for this reason an important part of our study design addresses the problem of categorizing the types of documentation referenced. The part of the Ye et al. study that is the most complementary to ours is their analysis of the purpose of links. However, because the study focuses on internal links, their classification does not include purposes that would be exclusive to resources outside of Stack Overflow itself. Their classification is also more abstract, with four categories of purpose (excluding the ``other'' category), whereas we analyze the link purposes at a finer granularity.

In addition to Stack Overflow, studies have also investigated linking practices in other context. Hata et al.~\cite{Hata2019} studied the role of links contained in source code comments in terms of prevalence, link targets, purposes, decay, and evolutionary aspects.
They report that links can be fragile since link targets change frequently or disappear. Links are also shared as part of code review. Jiang et al. contribute a study of link sharing in review comments~\cite{LJL2019a}, reporting that roughly half the links they identified refer to resources outside the project. This observation further motivates our study in that the observations we make about information diffusion may also be applicable to contexts other than question and answer forums.

\section{Study Design}
\label{s:design}

To investigate \emph{how} and \emph{why} documentation resources are referenced in Stack Overflow threads, we conducted a mixed-methods study involving a qualitative analysis of 759 links from 742 different threads and a quantitative analysis using association rule mining and logistic regression models.

\subsection*{Research Questions}

The overall goal of the study is to discover the roles that links to documentation play in Stack Overflow threads and thus pave the way for a more systematic treatment of documentation references on Q\&A sites for software developers.
We split our research questions into two sub-questions:

\begin{description}
\item[\textbf{RQ1}] What is the \textbf{context} around documentation links in Stack Overflow threads? With this question we study \emph{how} links are provided.
\item[\textbf{RQ2}] What is the \textbf{purpose} that documentation links in Stack Overflow threads serve? With this question we study \emph{why} links are provided.
\end{description}

With these questions, our aim was to collect specific insights about linking practices on Stack Overflow, that can support actionable implications for authors and readers of Q\&A forums and for the development of technology based on the analysis of such forums.

Our first research question was motivated by the fact that Stack Overflow encourages users to provide \textbf{context} for links~\cite{SO2019a}, in particular by quoting external sources~\cite{SO2019b}.
We qualitatively analyzed whether users follow this advice (see Section~\ref{s:qualitative}), but we also built logistic regression models capturing different features of Stack Overflow posts to quantitatively analyze which of those features are related to the presence of documentation links (see Section~\ref{sec:model-building}).

As the examples in Section~\ref{s:examples} illustrate, links on Stack Overflow serve diverse \textbf{purposes}.
To conduct a structured analysis of those purposes, we first built a classifier that was able to identify links to the most frequently referenced documentation resources (see Section~\ref{s:sampling}).
Based on a stratified sample of documentation links identified using the classifier, all three authors independently coded the purpose of 759 links using a jointly developed coding guide (see Section~\ref{s:qualitative}).
We mined the resulting data for association rules between documentation resources and assigned purposes and then used our qualitative and quantitative results to corroborate five major findings about linking practices on Stack Overflow (see Section~\ref{s:results}).

\subsection*{Cases Studied: Regex and Android}

Because even a cursory inspection of Stack Overflow threads shows clear differences in the use of references to external documentation, we structured our research as a multi-case study of linking practices for two different domains: use of regular expressions in Java (\emph{Regex}), and Android development (\emph{Android}). We bounded our investigation to clearly-defined domains to support a richer analysis of linking practices in the context of the wider documentation ecosystem they integrate. We selected \emph{Regex} and \emph{Android} because they constituted two very different domains (library vs. framework, small vs. large, integrated in the programming language vs. third-party, theoretically vs. practically grounded), and because we were familiar with both technologies. The importance of this latter aspect is not to be underestimated as a contributor to the meaningfulness of qualitative data analysis.

\subsection*{Overview of the Research Process}

\begin{figure*}[ht]
\centering
\includegraphics[width=1\textwidth,  trim=0.0in 0.0in 0.0in 0.0in]{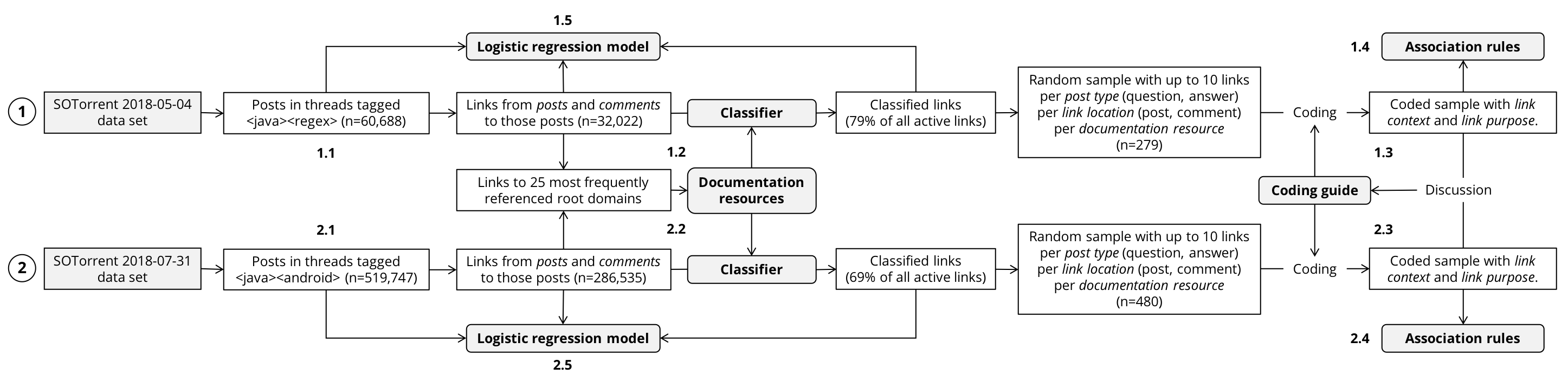} % left, bottom, right, top
\caption{Study process for both cases (1: \emph{Regex} and 2: \emph{Android}). The two cases were studied sequentially.}
\label{fig:study-design}
\vspace{-0.5\baselineskip}
\end{figure*}

Despite the ready availability of structured data from Stack Overflow, generating reliable insights about linking practices requires an extensive combination of analytical processing and manual inspection. Figure~\ref{fig:study-design} outlines the general process we followed. The research proceeded sequentially: we first completed an entire iteration for \emph{Regex} (referenced as number 1 on the figure), and then repeated the process for the second case (\emph{Android}), referenced as number 2.

In the following description, numbers refer to the step in the process overview (indicated after the period in Figure~\ref{fig:study-design}).

The first step was to retrieve all Stack Overflow threads related to each case ($N$.1). For this purpose we utilized the \emph{SOTorrent}~dataset~\cite{BDTD2018}.
For the \emph{Regex} case, we retrieved all threads with tags \C{java} and \C{regex}, and for the \emph{Android} case, the threads with tags \C{java} and \C{android}.
For each case, we used the most recent release at the time (2018-05-04 for the \emph{Regex} case~\cite{BD2018a} and 2018-07-31 for the \emph{Android} case~\cite{BD2018b}).

The second step was to process the links to determine what they were linking to, and to abstract the target of the links to one of a small set of \emph{documentation resource} categories (e.g., links to other Stack Overflow threads vs. links to API documentation).
We built a URL mapper to classify links to such documentation resources using the 25 most frequently referenced root domains for each case (Section~\ref{s:sampling} and $N$.2 in Figure~\ref{fig:study-design}).

The classification of links was necessary to create a stratified sample for detailed analysis, i.e., a sample guaranteed to contain links to all different types of resources. The third step was then to draw samples containing links to all identified documentation resources and qualitatively analyze their context and purpose (see Section~\ref{s:qualitative} and $N$.3 in Figure~\ref{fig:study-design}). This step involved extensive manual inspection and labeling of links in their context.

In step four, to investigate the motivation behind linking to documentation resources of a certain type, we used association rule mining~\cite{AIS1993} to investigate the relationship between resource type and purpose (see Section~\ref{sec:ar-mining} and $N$.4 in Figure~\ref{fig:study-design}).

Finally, we built logistic regression models to analyze which properties capturing the question context attract links to documentation resources in comments and answers (see Section~\ref{sec:model-building} and $N$.5 in Figure~\ref{fig:study-design}). In these models, we treat question features as independent variables and the presence of a link to a particular resource as dependent variable.

%\todo{@Christoph: Can we be a bit more specific here?}

\subsection*{Replication Package}

To support the complete replicability of this process and the verification of the results presented in this paper, we provide our coding guide, samples, and the analysis and data retrieval scripts as supplementary material~\cite{SupplementaryMaterial}.

\section{Link Classification and Sampling}
\label{s:sampling}

% We sampled threads for the regex case by filtering with the Stack Overflow tags tags \C{java} and \C{regex}, and for the \emph{Android} case with tags \C{java} and \C{android}. To retrieve the links from the posts and comments in those threads, we used the SOTorrent dataset~\cite{BDTD2018}.
% For each case, we used the most recent release at the time (2018-05-04 for the regex case~\cite{BD2018a} and 2018-07-31 for the \emph{Android} case~\cite{BD2018b}).

\begin{table}[tb]
\caption{Five most frequently referenced root domains and assigned documentation resources in \emph{Regex}; the second column lists the number of posts referring to the corresponding domain as well as the frequency relative to all posts containing links ($n_\text{posts} = 21,758$).}
\label{tab:java-regex-root-domains}
\centering
\begin{tabular}{lrl}
\toprule
\textbf{Domain} & \multicolumn{1}{c}{\textbf{\#Posts (\%)}} & \textbf{Resource Categories} \\
\midrule
stackoverflow.com & 5,120 (23.5\%) & {\scriptsize\emph{StackOverflow},}\\
& & {\scriptsize\emph{NotDocumentation}}\\
regex101.com & 4,439 (20.4\%) & {\scriptsize\emph{NotDocumentation}} \\
oracle.com & 4,316 (19.8\%) & {\scriptsize\emph{JavaAPI}, \emph{JavaReference},}\\
& & {\scriptsize\emph{OtherForum}} \\
ideone.com & 1,933 \phantom{1}(8.9\%) & {\scriptsize\emph{NotDocumentation}} \\
regular-expressions.info & 1,868 \phantom{1}(8.6\%) & {\scriptsize\emph{IndependentTutorial}} \\
\bottomrule
\end{tabular}
\end{table}

\begin{table}[tb]
\caption{Five most frequently referenced root domains and assigned documentation resources in \emph{Android}; the second column lists the number of posts referring to the corresponding domain as well as the frequency relative to all posts containing links ($n_\text{posts} = 177,784$).}
\label{tab:java-android-root-domains}
\centering
\begin{tabular}{lrl}
\toprule
\textbf{Domain} & \multicolumn{1}{c}{\textbf{\#Posts (\%)}} & \textbf{Resource Categories} \\
\midrule
stackoverflow.com & 57,461 (32.3\%) & {\scriptsize\emph{StackOverflow},} \\
& & {\scriptsize\emph{NotDocumentation}}\\
android.com & 42,199 (23.7\%) & {\scriptsize\emph{AndroidAPI}, \emph{AndroidReference}} \\
imgur.com & 22,339 (12.6\%) & {\scriptsize\emph{NotDocumentation}} \\
github.com & 18,259 (10.3\%) & {\scriptsize\emph{OtherReference},}\\
& & {\scriptsize\emph{NotDocumentation}} \\
google.com & 11,924 \phantom{1}(6.7\%) & {\scriptsize\emph{AndroidIssue}, \emph{AndroidReference},}\\
& & {\scriptsize\emph{OtherReference}, \emph{OtherForum}} \\
\bottomrule
\end{tabular}
\end{table}

\begin{table}
\caption{Assigned documentation resources for links in Stack Overflow posts and comments (NotDocumentation: links that our URL mapper classified as not pointing to a documentation resource, NotClassified: links that our URL mapper could not classify, DeadOrInvalid: links that were either unavailable or invalid).}
\label{tab:doc-resources}
\centering
\begin{tabular}{lr@{ }rr@{ }r}
\toprule
\textbf{Resource Category} & \multicolumn{2}{c}{\textbf{\#Links in \emph{Regex}}} & \multicolumn{2}{c}{\textbf{\#Links in \emph{Android}}} \\
\midrule
All & 35,022 & (100.0\%) & 286,535 & (100.0\%) \\
Classified & 25,917 & (74.0\%) & 185,857 & (64.9\%)\\
\hspace{1em}\textbf{Documentation} & \textbf{15,430} & (44.1\%) & \textbf{150,630} & (52.6\%) \\
\hspace{1em}NotDocumentation & 10,487 & (29.9\%) & 35,227 & (12.3\%) \\
NotClassified & 7,115 & (20.3\%) & 83,989 & (29.3\%) \\
InvalidOrDead & 1,990 & (5.7\%) & 16,689 & (5.8\%) \\
\midrule
\textbf{Documentation} & \textbf{15,430} & (100.0\%) & \textbf{150,630} & (100.0\%) \\
\hspace{1em}\emph{StackOverflow} & 5,656 & (36.7\%) & 64,610 & (42.9\%) \\
\hspace{1em}\emph{JavaAPI} & 5,093 & (33.0\%) & 7,403 & (4.9\%) \\
\hspace{1em}\emph{IndependentTutorial} & 2,419 & (15.7\%) & 6,600 & (4.4\%) \\
\hspace{1em}\emph{JavaReference} & 957 & (6.2\%) & 3,860 & (2.6\%) \\
\hspace{1em}\emph{Wikipedia} & 787 & (5.1\%) & 5,218 & (3.5\%) \\
\hspace{1em}\emph{OtherAPI} & 253 & (1.6\%) & 644 & (0.4\%) \\
\hspace{1em}\emph{OtherReference} & 262 & (1.7\%) & 6,514 & (4.3\%) \\
\hspace{1em}\emph{OtherForum} & 3 & (0.0\%) & 549 & (0.4\%) \\
\hspace{1em}\emph{AndroidAPI} & N/A & (0.0\%)  & 28,690 & (19.0\%) \\
\hspace{1em}\emph{AndroidReference} & N/A & (0.0\%) & 23,421 & (15.5\%) \\
\hspace{1em}\emph{AndroidIssue} & N/A & (0.0\%) & 1,301 & (0.9\%) \\
\hspace{1em}\emph{YouTube} & N/A & (0.0\%) & 1,820 & (1.2\%) \\
\bottomrule
\end{tabular}
\end{table}

% To be able to investigate the \emph{context} and \emph{purpose} of documentation links, we needed to extract links to \emph{documentation resources}.

Links on Stack Overflow may point to resources other than documentation, e.g., tools or images.
To be able to study links to documentation resources on Stack Overflow, we built a URL-based classifier that takes as input a link and outputs either one of 12 documentation resource categories that best describes the target of the link, or marks the link as \emph{NotDocumentation} (see Table~\ref{tab:doc-resources}).
Those 12 categories emerged during an iterative analysis of the most frequently linked domains.
We used the classifier to categorize all links in the two cases and then sampled links from each category of documentation links for our qualitative analysis.

\subsection*{Building the Classifier}

As mentioned above, we built the link categorization and corresponding classifier following a grounded, iterative approach.

First, we ranked all referenced \emph{root domains} according to the number of posts in which they were referenced (the root domain of \C{en.wikipedia.org}, for example, is \C{wikipedia.org}).
Starting with the most frequently referenced root domain, we inspected the extracted links and either decided that they form a new resource category or assigned them to an existing one.
Tables~\ref{tab:java-regex-root-domains} and \ref{tab:java-android-root-domains} show the five most frequently referenced root domains, meaning that those were the first five domains we derived resource categories from.

For each of the analyzed domains, we started by investigating the different paths that were linked from the Stack Overflow posts retrieved for the particular case.
For both cases, the most frequent link target was the platform itself.
Because such links are internal to platform, we created a dedicated documentation resource category \emph{StackOverflow}.
However, we soon realized that not all links to \C{stackoverflow.com} can be considered software documentation links, because the linked paths included user profiles (e.g., \C{/users/1974143}) or internal help pages (e.g., \C{/help/mcve
}).

Instead of excluding the paths that we did not consider documentation targets, we followed a whitelisting approach.
We first built regular expressions matching the paths of the domains that we identified as pointing to documentation resources (e.g., another Stack Overflow post).
After integrating those regular expressions in our link classifier, we executed the classification and analyzed the links to the current domain that had not been classified yet.
We then refined the regular expressions and repeated the process until all links to documentation resources were classified either as \textit{Documentation}, \textit{NotDocumentation}, or \textit{InvalidOrDead} (see also Table~\ref{tab:doc-resources}).
This process was performed by two authors who continuously discussed the emerging resource categories and associated regular expressions.
All decisions in the process were made unanimously.
The source code of the classifier, including the regular expressions for all documentation resources, is available on GitHub\footnote{\url{https://github.com/sbaltes/condor}} and archived on Zenodo~\cite{BaltesRobillardOthers2019}.

To conclude the above example, for the \C{stackoverflow.com} root domain we decided to only match links to questions, answers, post revisions, and comments---but not links to user profiles or pages with tips on how to write questions and answers (see above).
To illustrate this classification approach, we briefly describe the path matching for this domain.
As mentioned above, we modeled internal links as a separate documentation resource.
The regular expressions for the corresponding \emph{StackOverflow} documentation resource all start with:

{\scriptsize
\begin{quote}
\begin{verbatim}
^https?://((www|pt|ru|es)\\.)?stackoverflow\\.com
\end{verbatim}
\end{quote}
}

This prefix is followed by expressions matching the different paths we determined to point to documentation resources:

{\scriptsize
\begin{quote}
\begin{verbatim}
/(a|q|questions)/[\\d]+.*
/revisions.*
/posts/\\d+/revisions.*
/posts/comments.*
\end{verbatim}
\end{quote}
}

All other paths for the root domain \C{stackoverflow.com} are automatically classified as \C{NotDocumentation}.
Root domains that we have not analyzed yet are automatically labeled as \C{NotClassified}, root domains that our tool determined to be invalid or dead are automatically classified as \C{InvalidOrDead}.
This allowed us to track our progress.
We continued with the next root domain once we could not find paths anymore that were incorrectly labeled as \C{NotDocumentation}. 

We repeated the classification process for the 25 most frequently referenced root domains in both samples, which enabled us to classify 78.5\% of all \textit{active links} in the \emph{Regex} sample and 68.9\% of all active links in the \emph{Android} sample.
The ratio of classified active links can be derived from the data in Table~\ref{tab:doc-resources} as follows: $\nicefrac{(\text{Documentation} + \text{NotDocumentation})}{(\text{All} - \text{Dead})}$.
Because we conducted our analysis of the \emph{Android} case after the \emph{Regex} case had been completed, the classifier for \emph{Android} links was built by extending the preliminary \emph{Regex} link classifier.
Note that, as a last step, we re-ran the final classifier for the \emph{Regex} case.

Table~\ref{tab:doc-resources} shows the documentation resources we extracted for both cases.
In the following, we briefly describe which kinds of documentation resources we assigned to the different categories together with exemplary links.

%\todo{Does it make sense to move the following into a table?}

\textbf{\emph{StackOverflow:}} This documentation resource consists of Stack Overflow questions, answers, post revisions, and comments (see details above).

\textbf{\emph{OtherForum:}} We used this category to capture links to non-Stack-Overflow forum posts or threads including certain subpages of \C{forums.oracle.com} and \C{groups.google.com}.

\textbf{\emph{\{Java$\vert$Android$\vert$Other\}\{Reference$\vert$API\}:}}  The resource category \emph{JavaReference} represents official Java documentation except for the Java API documentation, which is represented by \emph{JavaAPI}. 
\emph{OtherReference}, \emph{AndroidReference}, \emph{OtherAPI}, and \emph{AndroidAPI} are analogously defined.
Examples for \emph{OtherAPI} include API documentation hosted on \C{jsoup.org}, \C{commons.apache.org}, and \C{developers.facebook.com}.
Examples for \emph{OtherReference} includes the \C{cookbook} subpages of  \C{jsoup.org}, certain reports on \C{unicode.org}, and different GitHub Pages.\footnote{\url{https://pages.github.com/}}

\textbf{\emph{AndroidIssue:}} Since Android issue descriptions were quite frequently referenced in the \emph{Android} case, we created a dedicated category for them.
Those links typically point to subpages of \C{issuetracker.google.com} or \C{code.google.com/p/android/issues/}.

\textbf{\emph{IndependentTutorial:}} Links in this category point to independent tutorials.
By `independent', we mean tutorials not provided by authoritative entities such as Oracle for \emph{Java} or Google for \emph{Android}.
Examples include \C{tutorialspoint.com}, \C{mkyong.com}, and \C{rexegg.com}.

\textbf{\emph{Wikipedia:}} We assigned links to Wikipedia pages in various languages to this category.

\textbf{\emph{YouTube:}} Especially in the \emph{Android} case, Stack Overflow users frequently referenced YouTube videos. We assigned such links to this category.

Section~\ref{s:quantitative} provides further examples for specific documentation resources, together with associated purposes we identified.
Table~\ref{tab:java-regex-root-domains} lists the five most frequently referenced root domains for \emph{Regex}, together with the number of links to those domains and the assigned resource categories.
Table~\ref{tab:java-android-root-domains} lists this information for \emph{Android}.

\subsection*{Sampling}

Because of the high effort involved in reviewing each link manually, we produced a sample of links to documentation resources for the qualitative analysis. We randomly sampled (up to) 40 links per documentation resource: We selected 20 links from questions (10 from question posts and 10 from question comments) and 20 links from answers (10 from answer posts and 10 from answer comments). Because some documentation resources had insufficient links to fulfill all of those selection constraints, the \emph{Regex} sample contained only 279 links (and not $8 \cdot 40 = 320$). The \emph{Android} sample contained $12 \cdot 40 = 480$ links, because we added four additional documentation resources that were only exhibited in that domain (see Table~\ref{tab:doc-resources}).
Section~\ref{s:threats} discusses implications of this sampling approach.

\section{Qualitative Analysis}
\label{s:qualitative}

We qualitatively analyzed all links in our samples to build a first layer of interpretation for linking practices. Following our research questions, we organized the coding~\cite{C2014} along two dimensions, \textit{context} and \textit{purpose}. For analyzing the context, much information is already available directly in the posts (e.g., the text surrounding the links). For context, we designed the coding task to complement this information with insights that are impossible to extract automatically, namely, whether the text in the context includes a quote or a summary of the link target---or whether the link is provided without any context. For purpose, we were interested in producing an abstraction of the purpose of the link as it would appear to a third party who read the corresponding thread.

\subsection*{Development of the Coding Guide}

\begin{figure}
\centering
\includegraphics[width=\columnwidth,  trim=0.0in -0.2in 0.0in 0.0in]{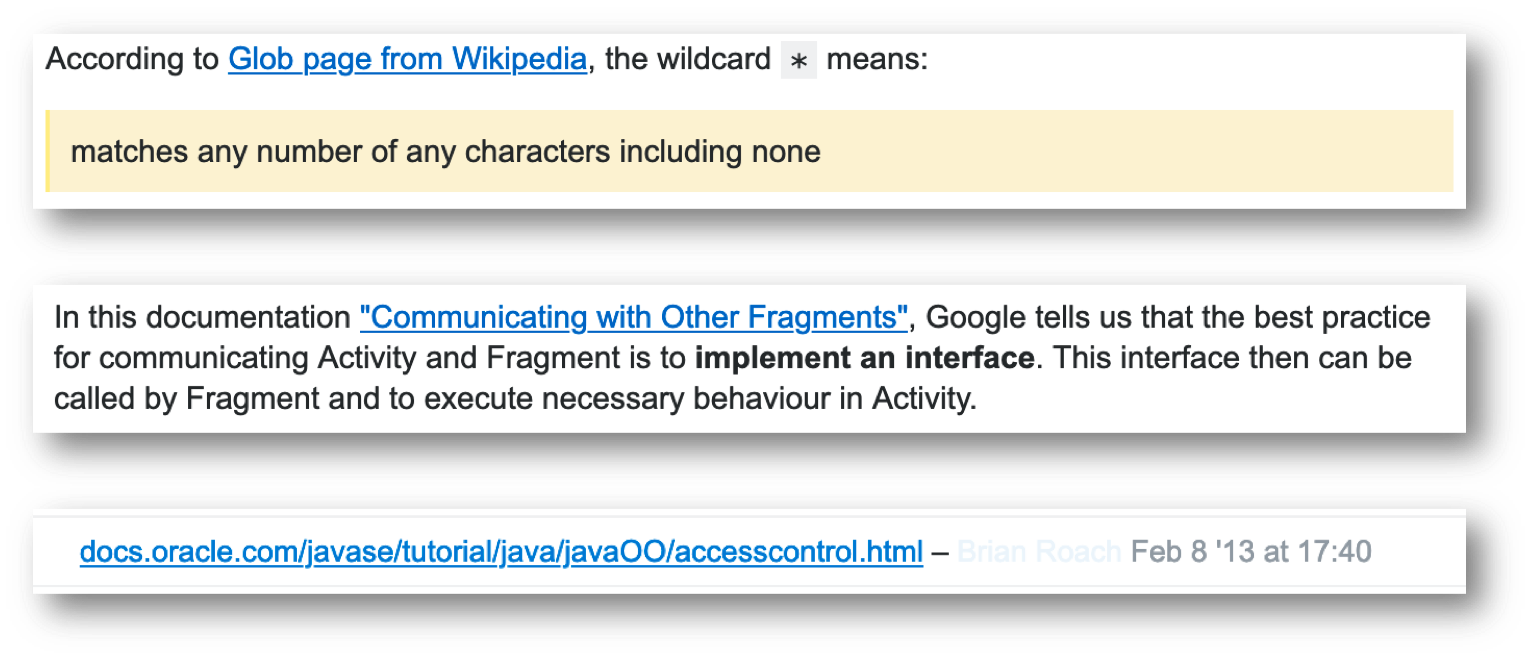} % left, bottom, right, top
\caption{Examples for context codes \textsc{Quote} (top), \textsc{Summary} (middle), and \textsc{LinkOnly} (bottom), taken from Stack Overflow questions \href{https://stackoverflow.com/q/37724969}{37724969}, \href{https://stackoverflow.com/q/28949786}{28949786}, and comment \href{https://stackoverflow.com/q/14778196\#comment20689962_14778196}{20689962}.}
\label{fig:context-codes}
\end{figure}

We developed a coding guide by considering the context and purpose dimensions separately.

\subsubsection*{Context}

For the context, creating the coding guide amounted to agreeing on what constituted a quote, a summary, and a link without context. 
The task was thus to indicate, for each link in the sample, \C{true} or \C{false} as values for the attributes \textsc{Quote} and \textsc{Summary}.
The attribute \textsc{Quote} indicates the presence of non-trivial content that has been copied without modification from the linked documentation resource into the Stack Overflow post or comment, the attribute \textsc{Summary} indicates that the Stack Overflow author provided at least one key insight from the linked documentation resource in their own words.
The third context code \textsc{LinkOnly} was assigned in case only the URL was provided (including anchor text) without any additional information surrounding it.
Note that while the codes \textsc{Quote}  and \textsc{Summary} can be assigned independent of the purpose codes, \textsc{LinkOnly} makes deriving a purpose impossible because no context is provided.
Therefore we modeled the former two as independent binary codes, as outlined above (see also Tables~\ref{tab:codes-regex} and \ref{tab:codes-android}).
Figure~\ref{fig:context-codes} illustrates the difference between the three context codes we assigned.

\subsubsection*{Purpose}

The development of a reliable coding guide for a link's purpose was much more challenging, and required multiple iterations. In an  \emph{initial coding} phase, we built a coding guide using a subset of the links for \emph{Regex}. During the initial coding, all three authors coded 80 links in four tasks of 20 each, discussing emerging categories after completing each task, until a stable coding guide emerged.
Prior to starting with the \emph{Android} sample, all three authors coded 50 links and then discussed if changes to the coding guide were required, which only led to one minor addition.
Note that, while the codes are not mutually exclusive, the coders always assigned one code that they considered to most accurately describe the link purpose. % added in response to ICSE review: "It’s not ever really clear whether or not the categories are mutually exclusive"
Table~\ref{t:codes} lists the codes with a brief description. The full description can be found in the supplementary material.
The modification that was required for the \emph{Android} case was simply to add ``watching a video'' to the code \textsc{BackgroundReading}, because of the new documentation resource \emph{YouTube}.

\begin{table*}[t]
\caption{Code catalog for link context and purpose (summary).}
\centering
\rowcolors{2}{}{black!10}
\begin{tabularx}{\textwidth}{llX} \toprule
\textbf{Abbrev.} & \textbf{Code} & \textbf{Description (Excerpt)}\\ \midrule

\textsc{QUO} & \textsc{Quote} &
Presence of non-trivial content copied without further modification.\\

\textsc{SUM} & \textsc{Summary} &
At least one key insight from the linked resource is provided in paraphrased from.\\

\textsc{ONL} & \textsc{LinkOnly} &
Link that only contains the URL (and anchor text) without any additional information surrounding it.\\

\midrule

\textsc{ATT} & \textsc{Attribution} &
Link to a resource simply to credit the source for material taken verbatim.\\

\textsc{AWA} & \textsc{Awareness} &
Link intended to make readers aware that a certain resources exists, or provide information about the nature of its content, without necessarily endorsing it.\\

\textsc{BGR} & \textsc{BackgroundReading} &
Link to a resource that a user thinks other users should read or watch to get better general knowledge of the topic related to the thread.\\

\textsc{CPT} & \textsc{Concept} &
Link to a resource that contains a general description of a concept that the reader should know about.\\

\textsc{CST} & \textsc{Consulted} &
Link to documentation to indicate that it was consulted prior to posting.\\

\textsc{LMN} & \textsc{LinkedMention} &
Link to the element-level (class, method, field) Javadocs of an API element that is mentioned as part of the text, without more specific indication for the purpose of the link.\\

\textsc{RCM} & \textsc{Recommendation} &
Link to resources that are landing pages for tools, libraries, API elements, or algorithms, for the purpose of recommending these.\\

\textsc{REF} & \textsc{Reference} &
Links to a resource to indicate the source of knowledge for an explicit claim, statement, or information conveyed in the post.\\

\textsc{OTH} & \textsc{Other} &
Link whose purpose is other than can be captured by other codes, unclear, or unknown.\\
\bottomrule

\end{tabularx}
\label{t:codes}
\end{table*}

\subsection*{Coding Process}

We used the coding guide in a \emph{focused coding} phase to go over all links in the sample and code them according to the guide, which we provide as supplementary material. All three authors used the coding guide to independently code the links by opening the Stack Overflow thread in a web browser, locating the link, and analyzing the surrounding context.

We coded the links in sets of up to 100 links, computing inter-rater agreement and discussing results after each set to ensure there were no major divergences or misunderstandings of the coding guide. To measure our inter-rater agreement, we calculated a three-way Cohen's kappa ($\kappa$)~\cite{C1960} for each set. Table~\ref{tab:cohen} presents the agreement data.

% For the regex sample, the $\kappa$ values were $0.61$ (100), $0.65$ (100), and $0.77$ (79).
% For the \emph{Android} sample, the values were $0.71$ (50), $0.70$ (100), $0.64$ (100), $0.74$ (100), $0.72$ (100), $0.80$ (30).

\begin{table}[h]
\caption{Inter-rater agreement for link purpose coding, with number of items in the set (\#) and corresponding $\kappa$ value.}
\label{tab:cohen}
\centering
\begin{tabular}{l|rrr|rrrrrr}
\toprule
& \multicolumn{3}{c}{\textbf{\emph{Regex}}} & \multicolumn{6}{|c}{\textbf{\emph{Android}}} \\ \midrule
\# & 100 & 100 & 79 & 50 & 100 & 100 & 100 & 100 & 30 \\
$\kappa$ & 0.61 & 0.65 & 0.77 & 0.71 & 0.70 & 0.64 & 0.74 & 0.72 & 0.80\\
\bottomrule
\end{tabular}
\end{table}

The task of identifying the \emph{purpose} of a link turns out to be very challenging. In some cases, the purpose can be ambiguous or opaque. The difficulty of the task is reflected in the kappa values. Although they increase towards the end as we became more proficient, values in the 0.65-0.80 range, although usable, are indicative of a non-negligible amount of residual flexibility of interpretation.

The difficulty of the coding task is the reason we opted for the unusual and very labor intensive practice of coding every single item in our data set in triplicate. This decision significantly mitigates the threats of bias in the coding task, since we were able to systematically detect links with ambiguous purpose and resolve disagreements by applying the following formal process:
After each coding iteration, we merged the purpose and \textsc{LinkOnly} codes by selecting the code which at least two investigators used (majority vote), and assigned the code \textsc{Other} if there was no agreement, which happened for 14 \emph{Regex} links (5\%) and for 13 \emph{Android} links (2.7\%).
The binary codes capturing the link \emph{context} were assigned a value of \C{true} if at least two investigators considered the link to be accompanied by a \textsc{Quote} or \textsc{Summary} respectively.

\subsection*{Final Coding}

Tables~\ref{tab:codes-regex} and \ref{tab:codes-android} show the frequency of each code per documentation resource for both cases.
Examples for the three context codes can be found in Figure~\ref{fig:context-codes}.
Section~\ref{s:quantitative} presents examples for the purpose codes together with related developer resources.

While our URL mapper was able to detect most invalid or dead links, we still noticed some broken links in the samples (coded as N/A). We also coded links as N/A if they were not rendered on Stack Overflow's website, but present in the Markdown source of the posts or comments, which we used to extract the links from.

\begin{table*}
\caption{Documentation resources and corresponding codes (purpose and context) for \emph{Regex} case.}
\label{tab:codes-regex}
\centering
\begin{tabular}{lrrrrrrrrrrrrrr}
\toprule
 & \textsc{ATT} & \textsc{AWA} & \textsc{BGR} & \textsc{CPT} & \textsc{CST} & \textsc{RCM} & \textsc{REF} & \textsc{LMN} & \textsc{ONL} & \textsc{OTH} & N/A & Total & \textsc{Quote} & \textsc{Summary} \\
\midrule
\textit{StackOverflow} & 2 & 16 & 0 & 0 & 5 & 0 & 3 & 0 & 2 & 12 & 0 & 40 & 3 & 6 \\
\textit{JavaAPI} & 2 & 4 & 0 & 0 & 5 & 5 & 7 &12 & 1 & 3 & 1 & 40 & 7 & 10 \\
\textit{IndependentTut.} & 1 & 2 & 9 & 9 & 6 & 0 & 8 & 0 & 0 & 4 & 1 & 40 & 3 & 5 \\
\textit{JavaReference} & 2 & 4 & 12 & 3 & 6 & 2 & 6 & 1 & 1 & 3 & 0 & 40 & 3 & 7 \\
\textit{Wikipedia} & 1 & 1 & 2 & 22 & 1 & 3 & 3 & 0 & 1 & 4 & 2 & 40 & 1 & 9 \\
\textit{OtherAPI} & 1 & 6 & 0 & 0 & 3 & 17 & 0 & 8 & 0 & 4 & 1 & 40 & 2 & 6 \\
\textit{OtherReference} & 1 & 13 & 2 & 3 & 2 & 3 & 3 & 0 & 0 & 7 & 2 & 36 & 2 & 7 \\
\textit{OtherForum} & 0 & 1 & 0 & 0 & 0 & 0 & 0 & 0 & 0 & 1 & 1 & 3 & 0 & 0 \\
\midrule
Total & 10 & 47 & 25 & 37 & 28 & 30 & 30 & 21 & 5 & 38 & 8 & \textbf{279} & 21 & 50 \\
\bottomrule
\end{tabular}
\end{table*}

\begin{table*}
\caption{Documentation resources and corresponding codes (purpose and context) for \emph{Android} case.}
\label{tab:codes-android}
\centering
\begin{tabular}{lrrrrrrrrrrrrrr}
\toprule
 & \textsc{ATT} & \textsc{AWA} & \textsc{BGR} & \textsc{CPT} & \textsc{CST} & \textsc{RCM} & \textsc{REF} & \textsc{LMN} & \textsc{ONL} & \textsc{OTH} & N/A & Total & \textsc{Quote} & \textsc{Summary} \\
 \midrule
\textit{StackOverflow} & 1 & 23 & 0  & 0 & 4  & 0 & 3 & 0 & 2 & 6 & 1 & 40 & 0 & 3 \\
\textit{JavaAPI} & 3 & 4 & 0 & 1 & 3 & 14 & 3 & 11 & 0 & 1 & 0 & 40 & 3 & 7 \\
\textit{IndependentTut.} & 0 & 16 & 4 & 2 & 10 & 1 & 1 & 0 & 3 & 2 & 1 & 40 & 0 & 1 \\
\textit{JavaReference}  & 3 & 6 & 15 & 3 & 5 & 0 & 4 & 1 & 2 & 1 & 0 & 40 & 5 & 5 \\
\textit{Wikipedia} & 1 & 7 & 2 & 22 & 0 & 1 & 1 & 0 & 0 & 6 & 0 & 40 & 2 & 4 \\
\textit{OtherAPI} & 0 & 7 & 1 & 0 & 5 & 11 & 5 & 7 & 0 & 4 & 0 & 40 & 2 & 3 \\
\textit{OtherReference} & 1 & 12 & 0 & 2 & 6 & 4 & 7 & 0 & 2 & 4 & 2 & 40 & 2 & 8 \\
\textit{OtherForum} & 1 & 22 & 0 & 0 & 9 & 0 & 2 & 0 & 0 & 6 & 0 & 40 & 1 & 3 \\
\midrule
\textit{AndroidAPI} & 2 & 8 & 2 & 0 & 3 & 10 & 8 & 7 & 0 & 0 & 0 & 40 & 4 & 5 \\
\textit{AndroidReference} & 3 & 11 & 11 & 0 & 4 & 4 & 2  & 1 & 1 & 2 & 1 & 40 & 3 & 5 \\
\textit{AndroidIssue} & 2 & 20 & 0 & 0 & 3 & 0 & 6 & 0 & 0 & 9 & 0 & 40 & 1 & 6 \\
\textit{YouTube} & 0 & 13 & 9 & 0 & 9 & 0 & 0 & 0 & 2 & 7 & 0 & 40 & 0 & 2 \\
\midrule
Total & 17 & 149 & 44 & 30 & 61 & 45 & 42 & 27 & 12 & 48 & 5 & \textbf{480} & 15 & 34 \\
\bottomrule
\end{tabular}
\end{table*}

\section{Quantitative Analysis}
\label{s:quantitative}

The qualitative analysis provides the foundation that enabled three quantitative analyses to better understand linking practices:

\begin{enumerate}
\item A systematic comparison of code distributions between our two cases, to relate differences to their context.
\item The mining of \emph{association rules} to detect correspondences between a resource type and a link purpose.
\item The building of logistic regression models, using question features as independent variables and presence of a link to a particular resource as dependent variable, to determine the characteristics of a Stack Overflow question that are related to the features of documentation links in an answer or a comment.
\end{enumerate}

\subsection*{Code Frequency Comparison}
\label{sec:code-frequency}

\begin{figure}
\centering
\includegraphics[width=0.95\columnwidth,  trim=0.0in 0.4in 0.0in 0.0in]{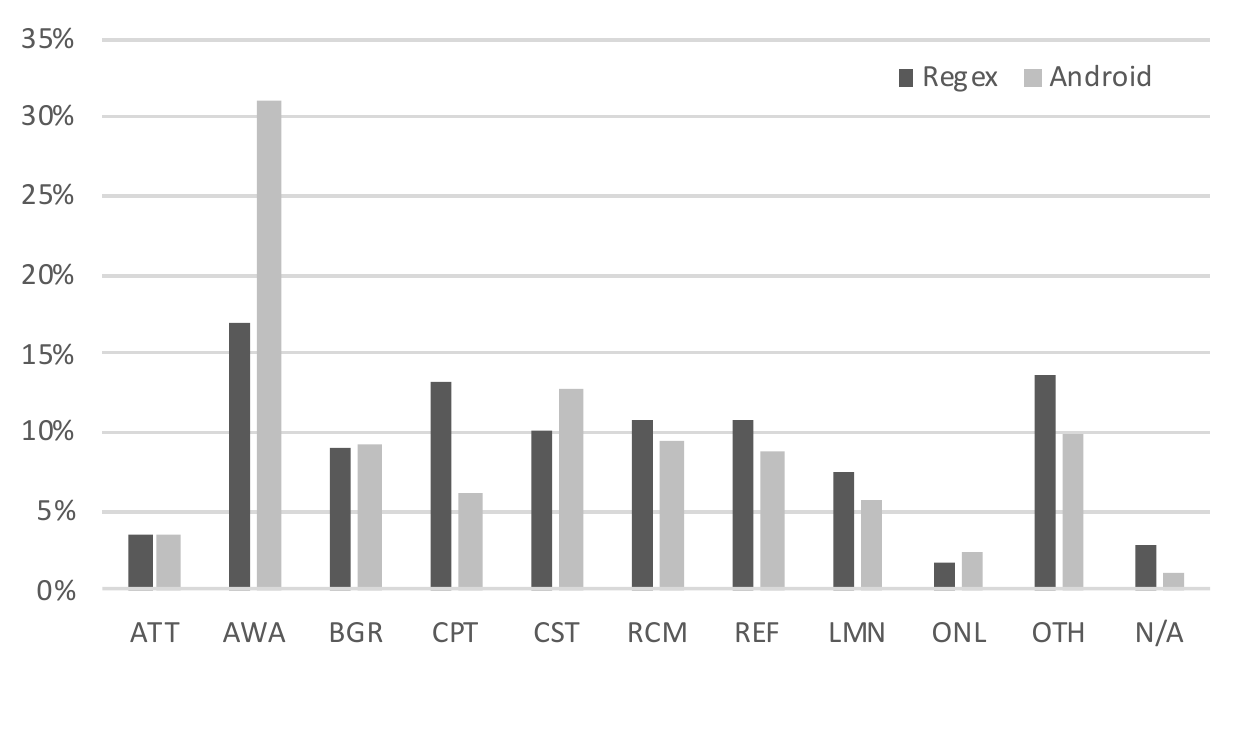} % left, bottom, right, top
\caption{Relative frequency of the assigned link purpose codes for both cases ($n_\text{regex} = 279$, $\;n_\text{android}=480$).}
\label{fig:code-frequencies}
\end{figure}

Figure~\ref{fig:code-frequencies} shows the relative frequency of the purpose codes we assigned.\footnote{Our use of stratified random sampling precludes the calculation of confidence intervals, which rely on an assumption of simple random sampling. As stated in the main text, the figure thus documents the code we assigned, without the implication that they would generalize to a population. This is consistent with our research goal and use of a case study method, whereby we sought to understand the phenomenon of link sharing as broadly as possible for two specific topics, as opposed to drawing implications for an entire dataset.} The bar charts reveals two major differences: in our sample, the code \textsc{Awareness} was about twice as common in the \emph{Android} case than in the \emph{Regex} case (31.0\% vs. 16.8\%). The reverse was true for the code \textsc{Concept}, which was about twice as common in the \emph{Regex} case (13.3\% vs. 6.3\%). Both difference were significant according to a two-tailed Fisher's exact test~\cite{F1922} with a significance level of $\alpha=0.01$.\footnote{The p-values were $0.0001$ for the \textsc{Awareness} frequency difference and $0.0014$ for the \textsc{Concept} frequency difference.}

Both of these differences can be directly linked to salient aspects of the technological environment of the cases analyzed. The sample for the \emph{Regex} case exhibits twice as many \textsc{Concept}-related links, which can be explained by the theoretical nature of the domain. The links we coded are to concepts such as \emph{context-free grammar} and \emph{regular language}. As for \emph{Android}, the extensive use of links for \textsc{Awareness} purposes can be explained by the huge size of this technology ecosystem, where many users end up posting answers and comments simply to point out relevant resources to each other.

\subsection*{Association Rule Mining}
\label{sec:ar-mining}

To distill the main motivation behind linking to documentation resources of a certain type, we mined association rules between resources and assigned purpose codes.

\subsubsection*{Method}

We first transformed the documentation resource categories as well as the purpose and \textsc{LinkOnly} codes into binary properties of the links, added the \textsc{Quote} and \textsc{Summary} codes, and then applied the \emph{apriori} algorithm~\cite{AS1994} as implemented in the R package \C{arules}\footnote{\url{https://cran.r-project.org/web/packages/arules/index.html}} to retrieve binary rules.

\begin{table}
\caption{Binary association rules between documentation resource type and purpose/context codes in the \emph{Regex} sample.}
\label{tab:ar-regex}
\centering
\begin{tabular}{p{5.5em}p{8.3em}llrrrr}
\toprule
\multicolumn{1}{c}{\textbf{LHS}} & \multicolumn{1}{c}{\textbf{RHS}} & \textbf{Supp} & \textbf{Conf} & \textbf{Lift} & \multicolumn{1}{c}{\textbf{n}} \\
\midrule
\scriptsize\textit{Wikipedia} & \scriptsize$\rightarrow$ \textsc{Concept} & 0.08 & 0.58 & 4.22 & 22 \\
\scriptsize\textit{OtherAPI} & \scriptsize$\rightarrow$ \textsc{Recomm.} & 0.06 & 0.45 & 3.92 & 17 \\
\scriptsize\textit{StackOverflow} & \scriptsize$\rightarrow$ \textsc{Awareness} & 0.05 & 0.40 & 2.30 & 16 \\
\scriptsize\textit{OtherReference} & \scriptsize$\rightarrow$ \textsc{Awareness} & 0.05 & 0.38 & 2.20 & 13 \\
\scriptsize\textit{JavaAPI} & \scriptsize$\rightarrow$ \textsc{LinkedMention} & 0.04 & 0.31 & 3.96 & 12 \\
\scriptsize\textit{JavaReference} & \scriptsize$\rightarrow$ \textsc{BackgroundR.} & 0.04 & 0.30 & 3.24 & 12 \\
%\scriptsize\textit{StackOverflow} & \scriptsize$\rightarrow$ \textsc{Other} & 0.04 & 0.30 & 2.19 & 12 \\
\midrule
\scriptsize\textit{Attribution} & \scriptsize$\rightarrow$ \textsc{Quote} & 0.04 & 1.00 & 12.85 & 10 \\
\scriptsize\textit{Reference} & \scriptsize$\rightarrow$ \textsc{Summary} & 0.08 & 0.73 & 3.96 & 22 \\
\bottomrule
\end{tabular}
\end{table}

\begin{table}
\caption{Binary association rules between documentation resource type and purpose/context in the \emph{Android} sample; rules only present in this sample are highlighted with a gray background.}
\label{tab:ar-android}
\centering
\begin{tabular}{p{5.5em}p{8.3em}llrrrr}
\toprule
\multicolumn{1}{c}{\textbf{LHS}} & \multicolumn{1}{c}{\textbf{RHS}} & \textbf{Supp} & \textbf{Conf} & \textbf{Lift} & \multicolumn{1}{c}{\textbf{n}} \\
\midrule
\scriptsize\textit{StackOverflow} & \scriptsize$\rightarrow$ \textsc{Awareness} & 0.05 & 0.59 & 1.9 & 23 \\
\scriptsize\textit{Wikipedia} & \scriptsize$\rightarrow$ \textsc{Concept} & 0.05 & 0.56 & 8.7 & 22 \\
\graybg\scriptsize\textit{OtherForum} & \graybg\scriptsize$\rightarrow$ \textsc{Awareness} & 0.05 & 0.55 & 1.8 & 22 \\
\graybg\scriptsize\textit{AndroidIssue} & \graybg\scriptsize$\rightarrow$ \textsc{Awareness} & 0.04 & 0.50 & 1.6 & 20 \\
\graybg\scriptsize\textit{IndependentTut.} & \graybg\scriptsize$\rightarrow$ \textsc{Awareness} & 0.03 & 0.41 & 1.3 & 16 \\
\scriptsize\textit{JavaReference} & \scriptsize$\rightarrow$ \textsc{BackgroundR.} & 0.03 & 0.38 & 4.0 & 15 \\
\graybg\scriptsize\textit{JavaAPI} & \graybg\scriptsize$\rightarrow$ \textsc{Recomm.} & 0.03 & 0.35 & 3.7 & 14 \\
\graybg\scriptsize\textit{Youtube} & \graybg\scriptsize$\rightarrow$ \textsc{Awareness} & 0.03  & 0.33  & 1.0 & 13 \\
\scriptsize\textit{OtherReference} & \scriptsize$\rightarrow$ \textsc{Awareness} & 0.03  & 0.32  & 1.0 & 12 \\
\graybg\scriptsize\textit{AndroidReference} & \graybg\scriptsize$\rightarrow$ \textsc{BackgroundR.} & 0.02  & 0.28  & 3.0 & 11 \\
\graybg\scriptsize\textit{AndroidReference} & \graybg\scriptsize$\rightarrow$ \textsc{Awareness} & 0.02  & 0.28  & 0.9 & 11 \\
\scriptsize\textit{JavaAPI} & \scriptsize$\rightarrow$ \textsc{LinkedMention} & 0.02 & 0.28 & 4.8 & 11 \\
\scriptsize\textit{OtherAPI} & \scriptsize$\rightarrow$ \textsc{Recomm.} & 0.02 & 0.28 & 2.9 & 11 \\
\graybg\scriptsize\textit{IndependentTut.} & \graybg\scriptsize$\rightarrow$ \textsc{Consulted} & 0.02  & 0.26 & 2.0 & 10 \\
\graybg\scriptsize\textit{AndroidAPI} & \graybg\scriptsize$\rightarrow$ \textsc{Recomm.} & 0.02  & 0.25 & 2.6 & 10 \\
\midrule
\scriptsize\textit{Attribution} & \scriptsize$\rightarrow$ \textsc{Quote} & 0.03 & 0.88 & 18.22 & 15 \\
\scriptsize\textit{Reference} & \scriptsize$\rightarrow$ \textsc{Summary} & 0.07 & 0.74 & 6.74 & 31 \\
\bottomrule
\end{tabular}
\end{table}

We note that the maximum support of a mined association rule is limited by the fact that we only sampled up to 40 links per documentation resource.
The \emph{Regex} sample, for example, contained 279 links in total (see Table~\ref{tab:ar-regex}).
If a rule is true for all 40 links to one particular resource, the support would still only be $\nicefrac{40}{279} = 0.14$.
In our analysis, we considered rules with at least 10\% of the maximum possible support, which was $\nicefrac{0.14}{10} = 0.014$ for the regex sample and $\nicefrac{0.08}{10} = 0.008$ for the \emph{Android} sample. Moreover, we excluded rules with less than 25\% confidence, meaning that a rule must be true in at least 1 out of 4 cases, and we further excluded rules involving the code \textsc{Other}.

\subsubsection*{Results}

Tables~\ref{tab:ar-regex} and \ref{tab:ar-android} show the binary association rules between the documentation resource types and the purpose/context codes.
In the following, we discuss those rules and provide illustrating examples.

\subsubsection*{\textit{Wikipedia} $\rightarrow$ \textsc{Concept}}

The purpose \textsc{Concept} was clearly associated with the resource \emph{Wikipedia}, having the highest and second highest confidence in the two samples, respectively. A typical usage scenario was to mention a concept related to the question and then use the first mention of the concept as link anchor pointing to the corresponding page on Wikipedia:

\begin{quote}
\sf
\footnotesize
I think you're using * as if it's the \textbf{Kleene star}, not * as Java, JavaScript, \& co. interpret * in regexps.~\cite{SO6}
% regex sample row 10
\end{quote}

This observation provides a clear characterization of the extent to which Wikipedia is leveraged to avoid defining concepts. The observation directly corroborates that of Vincent et al.\cite{VJH2018a}, who found that ``on SO, Wikipedia supports answers in the form of links and quoted text. Answers often use technical terms or acronyms and include a Wikipedia link in lieu of defining these terms.''

\subsubsection*{\textit{Java-}/\textit{OtherAPI} $\rightarrow$ \textsc{Recommendation}}

A second dominant group of association rules are related to \textsc{Recommendations}, which often pointed directly to the API documentation of a recommended software component. This is represented by the rule \emph{OtherAPI} $\rightarrow$ \textsc{Recommendation} in the regex sample and \emph{JavaAPI}/\emph{OtherAPI} $\rightarrow$ \textsc{Recommendation} in the \emph{Android} sample.

\begin{quote}
\sf
\footnotesize
You could use \textbf{Apache Commons Lang} for that...~\cite{SO7}
 % There you have methods like \textbf{isNumeric} and \textbf{isAlphanumeric}~\cite{SO7}
% \emph{Android} sample row 342
\end{quote}

% The above example not only shows how users recommend to use certain software components, but also how they refer to API documentation using a \textsc{LinkedMention}.
%
%
% Please note that we only applied that code if no more specific code applied (like \textsc{Recommendation} in the above example).
% \textsc{LinkedMentions} were related to \emph{JavaAPI} resources in both samples.
% They were, for example, used in questions to refer to related components:
%
% \begin{quote}
% \sf
% \footnotesize
% \textbf{java.util.regex.Pattern} is almost good enough, but its javadoc points out how it differs from Perl.~\cite{SO8}
% % regex sample row 229
% \end{quote}

% \textsc{Awareness} was the most frequently applied code in both samples, which is also represented by its presence in many of the association rules.
% Most salient was the usage of links to \emph{StackOverflow} posts to make readers aware of related solutions and discussions:
%
% \begin{quote}
% \sf
% \footnotesize
% ...you can use the \textbf{NetworkStatsManager} to get the user's historical data usage to find the oldest date for data usage for that sim. Look at \textbf{this} question on how to do that. .~\cite{SO9}
% % \emph{Android} sample row 125
% \end{quote}
%
% Sometimes, those \textsc{Awareness} links also carried a negative undertone:
%
% \begin{quote}
% \sf
% \footnotesize
% See this answer:\\
% \textbf{https://stackoverflow.com/questions/3919274/snippet-creation-keystroke-shortcut-in-eclipse}\\
% Make sure to google problems you might have before asking on StackOverflow.~\cite{SO10}
% % \emph{Android} sample row 75
% \end{quote}

\subsubsection*{\textit{Java-/AndoidReference} $\rightarrow$ \textsc{BackgroundReading}}

A main use case of reference documentation was providing readers with pointers to resources for \textsc{BackgroundReading}. This relationship is also reproduced in the association rules we identified, since \emph{JavaReference} were associated with \textsc{BackgroundReading} in both samples.
Moreover, \emph{AndroidReference} was associated with this purpose in the second sample. An example for \textsc{BackgroundReading} is provided below:

\begin{quote}
\sf
\footnotesize
Instead of asking people to code your regular expressions for you, try reading the Java Regular Expressions Tutorial. \textbf{...docs.oracle.com/javase/tutorial/..}.~\cite{SO14}
% regex sample row 247
\end{quote}

The above example illustrates the difference between the codes \textsc{Recommendation} and \textsc{BackgroundReading}.
We used \textsc{Recommendation} to highlight that the authors' primary intention was to recommend a specific tool or library (like in the example).
\textsc{BackgroundReading}, on the other hand, indicates that the author recommends a certain resource describing background knowledge relevant for the topic of the particular thread (see also descriptions of the codes in Table~\ref{t:codes}).

\subsubsection*{\textit{StackOverflow} $\rightarrow$ \textsc{Awareness}}

Other rules for link purposes were not as insightful because they rather confirmed the definition of our codes than indicated a particular linking practice. For example, although \emph{StackOverflow} $\rightarrow$ \textsc{Awareness} was a strong rule for both cases, it is hardly surprising that people will link to a Stack Overflow post to make others aware of it.

% \begin{quote}
% \sf
% \footnotesize
% If you have not done it yet, have a look at \textbf{Greedy, Reluctant, and Possessive Quantifiers} section of the Java RegEx tutorial. It seems to have some relevance to your task at hand.~\cite{SO11}
% % regex sample row 247
% \end{quote}

\subsubsection*{\textsc{Quote/Summary/LinkOnly}}

Regarding the \emph{context} of links, we only identified two rules that were present in both samples: \textit{Attribution} $\rightarrow$ \textsc{Quote} and \textit{Reference} $\rightarrow$ \textsc{Summary}.
The former indicates an obvious relationship between content copied from external sources and the purpose of attributing that content.
The latter indicates that especially for reference documentation, Stack Overflow authors felt the need to summarize key insights instead of copying content as-is.

Overall, quoting content was not very common in the posts and comments we analyzed.
In the \emph{Regex} sample, 7.5\% of the links referred to content being quoted, in the \emph{Android} sample only 3.1\% (see Table~\ref{tab:codes-regex}).
The quoted content ranged from complete code snippets to small parts of the reference documentation.
% \begin{quote}
% \sf
% \footnotesize
% From the \textbf{javadoc}
% \emph{\\w \hspace{.5em} A word character: [a-zA-Z\_0-9]}\\
% As you see there is no space, no single quote and no dot...~\cite{SO12}
% % \emph{Android} sample row 111
% \end{quote}
Summarizing linked resources was more common than quoting (17.9\% in \emph{Regex} and 7.1\% in \emph{Android}).
However, there was neither a summary nor a quote for 203 \emph{Regex} (72.8\%) and 400 \emph{Android} links (83.3\%), which can become a problem once the links are dead.

% \todo{Do we need the concrete numbers here (how many links had neither quote nor summary)?}
% The only rule we identified for summary, \textit{Reference}~$\rightarrow$~\textsc{Summary}, had a support of 0.08 and 0.07 with a confidence of 0.73/0.74 and $n=22/31$.
% An example for reference documentation being summarized is provided below.
%
% \begin{quote}
% \sf
% \footnotesize
% Once this is done, you'd just need to push the relevant parts through the appropriate hash implementation, and check the signature using the standard javax.security interfaces (as explained here: \textbf{docs.oracle.com/javase/tutorial/security/apisign/...}) ~\cite{SO13}
% % \emph{Android} sample row 243
% \end{quote}

% \begin{normalbox}
% \textbf{Association Rules:} Many association rules applied both in the \emph{Regex} and the \emph{Android} case, with varying support and confidence.
% The associations between the resource \emph{Wikipedia} and the purpose \textsc{Concept} and between \emph{StackOverflow} and \textsc{Awareness} were among the three rules with the highest confidence in both cases.
% \end{normalbox}

\subsection*{Model Building}
\label{sec:model-building}

% An important part of the context of any documentation link posted on Stack Overflow is the question which started the thread. Therefore, we also investigate the relationship between properties of questions and the presence of documentation links. To investigate which characteristics of a Stack Overflow question might explain whether it will attract a documentation link in an answer or a comment, we built logistic regression models, using question features as independent variables. As indicated in Figure~\ref{fig:study-design}, we built separate models for questions about regular expressions and questions about \emph{Android}, and we also built separate models to explain the presence of different types of resources.

To investigate which properties of a Stack Overflow question might explain whether it will attract documentation links, we built separate logistic regression models for the \emph{Regex} and \emph{Android} cases.

\subsubsection*{Data Preparation}

For each of the two cases (\emph{Regex} and \emph{Android}), the input data for the model building were three samples, each containing 100 Stack Overflow threads:

\begin{itemize}
\item \textbf{\emph{Documentation links:}} One sample with threads that attracted links to documentation resources. To identify such threads, we relied on our previous classification and randomly selected 100 threads with at least one answer or comment containing a link classified as pointing to one of the documentation resources (see Table~\ref{tab:doc-resources}).
\item \textbf{\emph{Non-documentation links:}} One sample with threads that attracted links, but not to documentation resources. We randomly selected 100 threads with at least one answer or comment containing a non-classified or non-documentation link (see Table~\ref{tab:doc-resources}).
\item \textbf{\emph{No links:}} One sample with threads that did not attract links at all. To draw this sample, we utilized the \emph{SOTorrent} dataset and selected only threads without any links in answers and comments (no records in tables \C{PostVersionUrl} and \C{CommentUrl}).
\end{itemize}

Our data retrieval and sampling scripts are available as part of the supplementary material.
Two of the authors independently analyzed all 600 threads to verify that they are indeed a representative of the corresponding class.
In case we found contradicting evidence (e.g., a link to a documentation resource in one of the non-documentation samples), we excluded those threads and then sampled and analyzed replacements.

\subsubsection*{Non-documentation Resources}

In the course of analyzing the two non-documentation samples, we also coded the purposes of those links.
In the \emph{Regex} sample, the most common purposes of non-documentation links were referring to a (regex) \emph{tool} (46), \emph{source code} (19), or websites with \emph{posting recommendations}\footnote{Examples: \url{http://whathaveyoutried.com/}~or~\url{http://sscce.org/}} (16).
In the \emph{Android} sample, the most common purposes were linking \emph{source code} (28), an online \emph{tool} (22, e.g., JSON or XML validators), or an \emph{image file} (19, e.g., icons or screenshots).

\begin{table}
\centering
\caption{Features of Stack Overflow posts used as independent variables in the logistic regression models.}
\label{tab:features}
\begin{tabular}{ll}
\toprule
\textbf{Feature} & \textbf{Description} \\
\midrule
TitleLength & \# of characters in question title\\
TextBlockCount & \# of text blocks in question\\
CodeBlockCount & \# of code blocks in question \\
LineCountText & \# of lines of text in question\\
LineCountCode & \# of lines of code in question\\
LengthText & \# of characters formatted as text\\
LengthCode & \# of characters formatted as code\\
UserAgeWhenPosting & \# of days since account creation\\
UserReputation & reputation of user\\
LinkCount & \# of links in question\\
LinkSpecificity 	& 0: no link\\
				& 1: link to root domain\\
				& 2: path present\\
				& 3: path contains fragment identifier\\
\midrule
Tags & tags associated with the question\\
\hspace{.5em} \textit{(one feature per tag)} & \hspace{.5em} \textit{\emph{Regex}: 4 features, \emph{Android}: 3 features}\\
Words in title & the question title\\
\hspace{.5em} \textit{(one feature per word)} & \hspace{.5em} \textit{\emph{Regex}: 14 features, \emph{Android}: 2 features}\\
Words in body & all text in the question body\\
\hspace{.5em} \textit{(one feature per word)} & \hspace{.5em} \textit{\emph{Regex}: 86 features, \emph{Android}: 69 features}\\
Terms in code & all code in the question body\\
\hspace{.5em} \textit{(one feature per term)} & \hspace{.5em} \textit{\emph{Regex}: 23 features, \emph{Android}: 118 features}\\
\bottomrule
\end{tabular}
\end{table}

\subsubsection*{Features}

Table~\ref{tab:features} shows the features used as independent variables in the logistic regression models. The set of features consists of numeric features that can be extracted from the question, such as \textit{LengthText} or \textit{CodeBlockCount}. Note that we excluded features that would be unknown at the time when the question was posted, such as how many views the question attracted or its score.
We retrieved the data for the features from the \emph{SOTorrent} dataset, which contains the content of Stack Overflow posts separated into text and code blocks, collects links from posts and questions, and provides the metadata from the official Stack Overflow data dump.

For the textual features, shown in the bottom part of Table~\ref{tab:features}, we treated each token as a separate feature and used token frequency as feature values. We separated text into tokens using whitespace, and we removed stopwords\footnote{We used the ``Long Stopword List'' from \url{https://www.ranks.nl/stopwords}} and punctuation as well as special characters. All tokens were stemmed using the Porter stemming algorithm~\cite{porter1980algorithm}. We discarded features consisting of a single character such as a single digit, and we limited the set of features to tokens whose frequency in our dataset exceeded a minimum threshold. We used the goodness of fit (measured using McFadden's pseudo-$R^2$~\cite{mcfadden1973conditional}) to determine the best threshold for each dataset, resulting in a threshold of 15 for the \emph{Regex} dataset (McFadden's pseudo-$R^2$ = 0.549) and 22 for the \emph{Android} dataset (McFadden's pseudo-$R^2$ = 0.592). This led to a total of 138 features for the \emph{Regex} dataset and 203 features for the \emph{Android} dataset. Table~\ref{tab:features} shows the number of features resulting from each textual property.

The interpretation of logistic regression models may be misleading if the metrics that are used to construct them are correlated~\cite{tantithamthavorn2018experience}. As Table~\ref{tab:features} shows, some of our features are likely to be correlated, e.g., \textit{LineCountText} and \textit{LengthText}. To mitigate correlated metrics, we used AutoSpearman~\cite{jiarpakdee2018autospearman}, an automated metric selection approach based on correlation analyses, with a threshold of 0.7.
%AutoSpearman repeatedly selects one metric from a group of the most strongly correlated metrics that shares the least correlation with other metrics that are not in that group until no more uncorrelated metrics can be selected.

Following the advice of Tantithamthavorn and Hassan~\cite{tantithamthavorn2018experience}, we used ANOVA Type-II importance scores to interpret our logistic regression models after constructing them using the \texttt{glm} function in R.

\begin{table}
\centering
\caption{Most important features for explaining whether a Stack Overflow question related to regular expressions will attract a particular type of documentation link. All p-values $<$ 0.001.}
\label{tab:lr_resourcetypes_regex}
\begin{tabular}{lllrr}
\toprule
\textbf{Resource}              & \textbf{Origin} & \textbf{Feature} & \textbf{Coeff.} & \textbf{ANOVA} \\
\midrule
\emph{Wikipedia}         & Text     & pars & $+$13.1              & 9\%\\
                  & Text     & java & $+$10.2              & 8\%\\
                  & Text     & issu & $+$21.9              & 7\%\\
                  & Title    & pattern & $+$15.4           & 6\%\\
                  & Text     & problem & $-$16.9           & 5\%\\
\bottomrule
\end{tabular}
\end{table}

\begin{table}
\centering
\caption{Most important features for explaining whether a Stack Overflow question related to \emph{Android} will attract a particular type of documentation link. All p-values $<$ 0.001.}
\label{tab:lr_resourcetypes_android}
\begin{tabular}{lllrr}
\toprule
\textbf{Resource}              & \textbf{Origin} & \textbf{Feature} & \textbf{Coeff.} & \textbf{ANOVA} \\
\midrule

\emph{Wikipedia}         & Text     & devic & $+$15.7             & 17\%\\
                  & Text     & creat & $+$7.0             & 11\%\\
                  & Text     & user & $+$9.0              & 5\%\\
                  & Metadata & UserReputation & $+$0.0    & 4\%\\
                  & Text     & call & $-$11.4              & 3\%\\
\midrule
\emph{Stack}             & Text     & find & $+$46.8              & 8\%\\
\emph{Overflow}         & Code     & activ & $-$7.2             & 7\%\\
                  & Text     & click & $-$45.6             & 6\%\\
                  & Text     & call & $-$17.6              & 6\%\\
                  & Code     & edittext & $+$19.9          & 6\%\\
\midrule
\emph{JavaAPI}          & Text     & convert & $+$42.3           & 10\%\\
                  & Text     & phone & $-$76.8             & 6\%\\
                  & Text     & problem & $+$38.3           & 6\%\\
                  & Text     & string & $+$8.7            & 6\%\\
                  & Metadata & LineCountText & $+$4.4     & 5\%\\
\midrule
\emph{\emph{Android}}           & Metadata & UserReputation & $+$0.0    & 11\%\\
\emph{Reference}         & Code     & text & $-$29.7              & 6\%\\
                  & Code     & 065941702 & $+$6.8         & 5\%\\
                  & Code     & wsystemerr1249 & $+$7.1    & 5\%\\
                  & Code     & viewonclicklisten & $+$58.7 & 4\%\\
\bottomrule
\end{tabular}
\end{table}

\subsubsection*{Models For Documentation Resources}

We built logistic regression models for specific types of documentation resources. Note that we do not treat type of documentation resource as a categorical variable since posts can contain links to multiple different documentation resources.
While we did not have enough data to allow the construction of models for all types of resources, Tables~\ref{tab:lr_resourcetypes_regex} and \ref{tab:lr_resourcetypes_android} show the five most important features (as determined by the ANOVA Type-II test) for a subset of resource types for the \emph{Regex} and \emph{Android} datasets. %, respectively.
Table~\ref{tab:lr_resourcetypes_regex} indicates that \emph{Regex} questions about \textit{parsing} and \textit{patterns} are associated with a higher chance of attracting a link to \emph{Wikipedia}.
In contrast, questions about specific \textit{problems} are associated with a lower likelihood.
%Question 5375999~\cite{SO15} is exemplary of an information need about parsing being redirected to \emph{Wikipedia}:
%A Stack Overflow user asks about the speed differences between JDOM and regular expressions for XML parsing, mentioning the keyword \textit{parse} two times in the question body.
%In response, one of the answers recommends a SAX parser instead and provides a \textsc{Concept} link to the corresponding \emph{Wikipedia} article.\footnote{\url{http://en.wikipedia.org/wiki/Simple_API_for_XML}}
%As an example of a question about a specific problem and therefore less likely to attract a \emph{Wikipedia} link, question 5035500~\cite{SO16} did not attract any link---the question is about the specific problem of lookahead failing, and the user demonstrates their problem with a JUnit test, explicitly mentioning the keyword \textit{problem} in the question body.
%In response to this question, the accepted answer provides a code snippet containing an additional test.
For \emph{Android}, questions about \textit{devices} are associated with a higher chance of attracting \emph{Wikipedia} links while questions about \textit{converting} are associated with attracting links to the \emph{JavaAPI}.
%Question 3505356~\cite{SO17} is exemplary for a question about devices attracting a link to \emph{Wikipedia}:
%The user is looking for a ``list of \emph{Android} devices with specs'', and an \textsc{Awareness} link to the corresponding \emph{Wikipedia} page\footnote{\url{http://en.wikipedia.org/wiki/List_of_Android_devices}} is provided by the user answering the question.
%In contrast, question 25454064~\cite{SO18} from our \emph{Android} sample is a good example for a question about \textit{converting} attracting a link to the \emph{JavaAPI}:
%A user asks for help for sorting posts from different RSS feeds by their timestamps, mentioning the word \textit{convert} twice in the question body text. Handling date and time issues is such a common concern in Java or \emph{Android} applications that existing API documentation links can be used to answer the question.
%Indeed, the accepted answer provides an \textsc{Awareness} link to the documentation of the \texttt{compareTo} method of \texttt{java.util.Date}.
%%\footnote{\url{https://docs.oracle.com/javase/7/docs/api/java/util/Date.html\#compareTo\%28java.util.Date\%29}}
As shown in Table~\ref{tab:lr_resourcetypes_android}, links to the \textsc{AndroidReference} documentation are associated with questions asked by users with a higher reputation.
%The median user reputation of users asking questions which attract links to the \textsc{AndroidReference} documentation in our dataset used for the logistic regression analysis is 1063.5, while the corresponding median for the remaining questions is 86.
Interestingly, a manual inspection of the corresponding questions suggests that many of these high-reputation users are outsiders whose expertise is in areas other than \emph{Android}.
% (see discussion of reputation-expertise mismatch in Section~\ref{s:results}).
%One example is question 4280507~\cite{SO19} from our sample: the user asking the question is an expert with a Stack Overflow reputation of above 25,000 and \textit{iphone} and \textit{objective-c} as his top tags. Two of the answers point him at a specific page of the \emph{Android} reference documentation.

We provide an interpretation of these results as part of the discussion in the next section. Ultimately, such models could be used to recommend the inclusion of different types of links in Stack Overflow posts.

\section{Findings}
\label{s:results}

%\todo{R2: The summary of this section is quite confusing. It tries to sum up what's coming next, while also attempting at not spoiling too much. Ultimately, it ends up lacking clarity. I think it would read better moved into Conclusion and used as a summary of findings, once they have been properly discussed.}

Our systematic analysis of the \emph{context} (\textbf{RQ1}) and \emph{purpose} (\textbf{RQ2}) of documentation links led to five major findings about linking practices on Stack Overflow. In this section, we detail the evidence for each finding and discuss its main implication.

\begin{figure}
\centering
\includegraphics[width=\columnwidth,  trim=0.2in 0.2in 0.2in 0.2in]{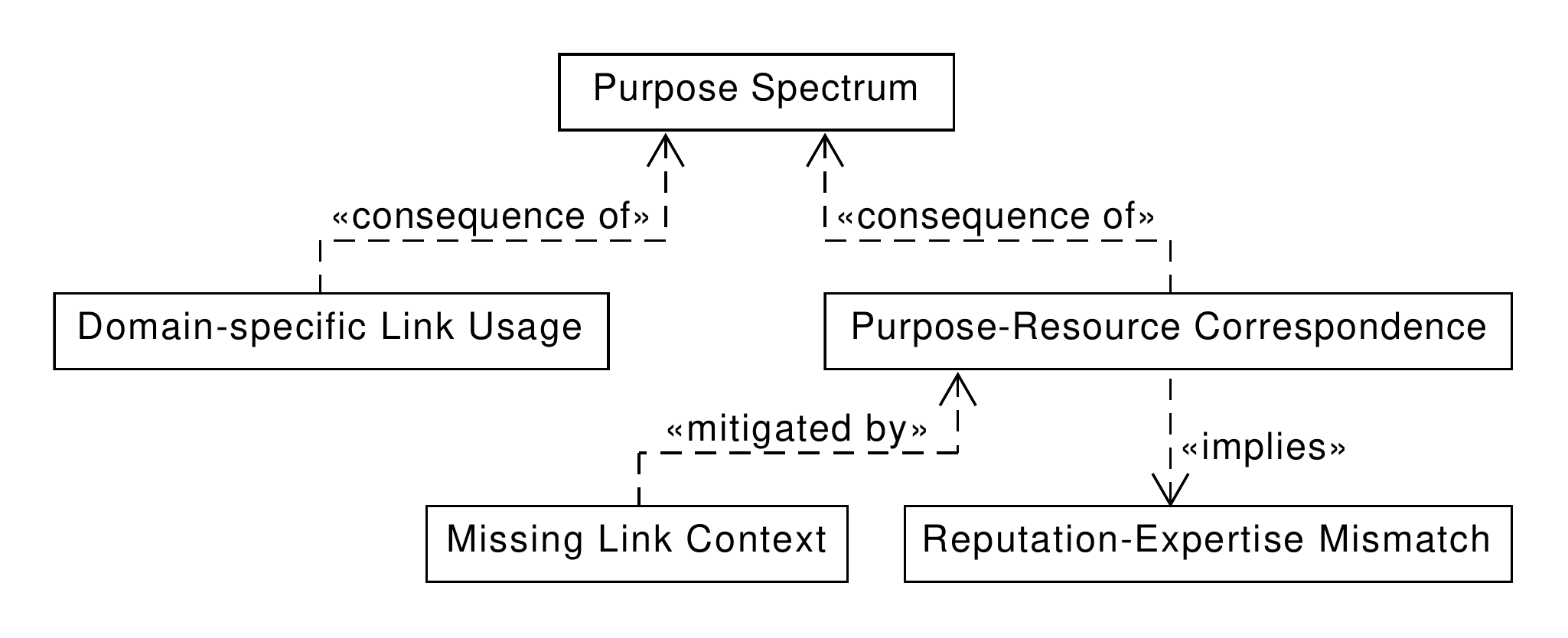} % left, bottom, right, top
\caption{Relationships between findings about linking practices on Stack Overflow.}
\label{fig:observations}
\end{figure}

\subsection*{Purpose Spectrum}

\begin{figure}
\centering
\includegraphics[width=0.8\columnwidth,  trim=0in 0in 0in 0in]{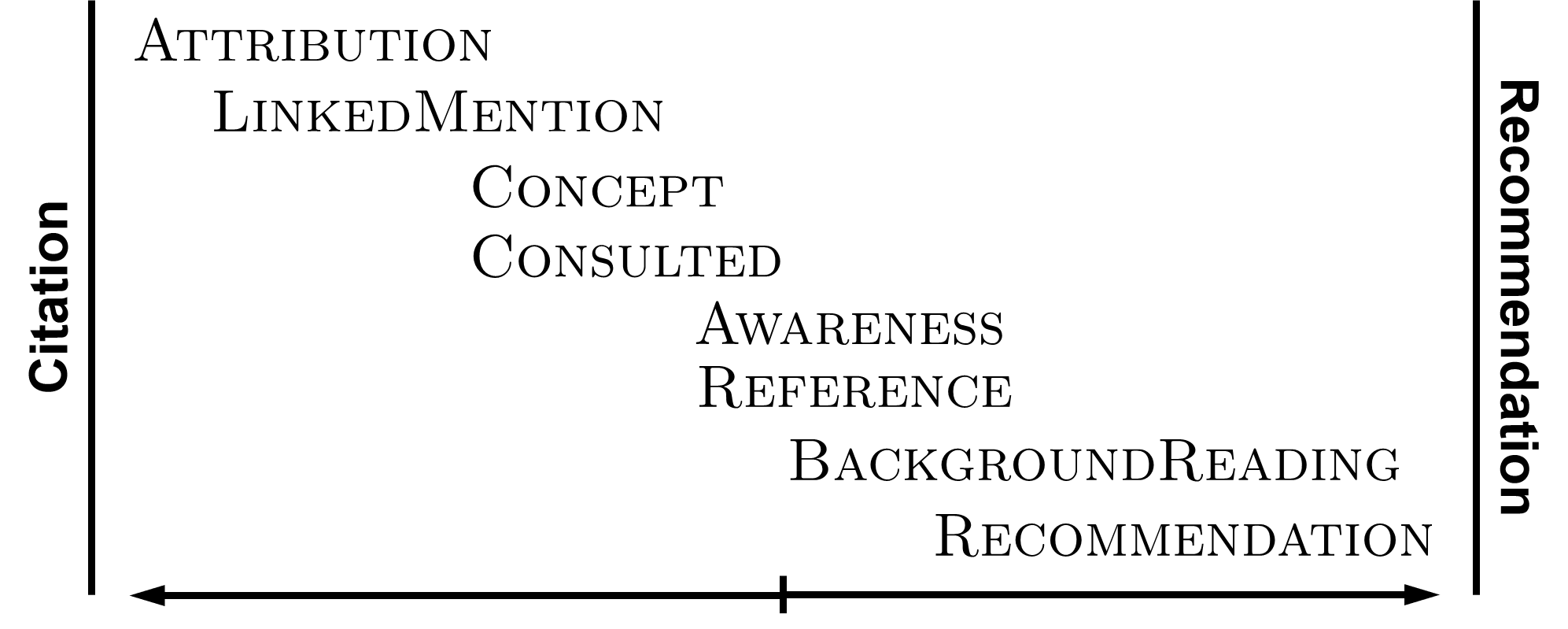} % left, bottom, right, top
\caption{Purpose codes arranged on the purpose spectrum from citation to recommendation.}
\label{fig:purpose-spectrum}
\end{figure}

Our qualitative analysis has shown that documentation links on Stack Overflow serve a variety of purposes. Figure~\ref{fig:code-frequencies} shows a rich diversity of purposes with eight of ten categories showing relative frequency above 5\%. Manually reviewing all the links (through the coding process) also showed that the different categories of link purposes can be positioned on a spectrum bounded by the concepts of \textit{Citation} and \textit{Recommendation}, where citations are not meant to be consulted whereas recommendations are explicit entreaties to follow the link. Figure~\ref{fig:purpose-spectrum} positions every link purpose category except for \textsc{Other} along this axis.

\textit{Citation} links include the ones labeled as \textsc{Attribution} and \textsc{LinkedMention}. The purpose of \textsc{Attribution} links is to credit the source of content copied into Stack Overflow, which can help users meet Stack Overflow's requirement for attribution~\cite{SD2018}. The purpose of the \textsc{LinkedMention} links is to uniquely identify a software artifact without the need to provide further context. Often, users add such \textsc{LinkedMention} references as inline links, which underlines their peripheral role:

\begin{quote}
\sf
\footnotesize
Is there a regex that would work with \textbf{String.split()} to break a String into contiguous characters...?\cite{SO20}
% Regex row 226: https://stackoverflow.com/q/13596454
\end{quote}

We place \textsc{Consulted} and \textsc{Concept} in the middle of the spectrum because they are open to interpretation. \textsc{Consulted} links are typically added for context, but in some cases this context is simply to show due diligence (closer to citation) and in some cases it is to point to an unclear document to be explained, e.g.,:

\begin{quote}
\sf
\footnotesize
I am trying to understand the regular expression in Solr and came across \textbf{this} Java doc where explains... having a hard time understanding what it really means. \cite{SO28}
% Regex row 192: https://stackoverflow.com/q/35762611
\end{quote}

As for \textsc{Concept} links, they are useful for readers who want to learn more about a mentioned concept, but they are usually also peripheral to the actual content of the post or comment (reproduced from Section~\ref{s:quantitative}).

\begin{quote}
\sf
\footnotesize
I think you're using * as if it's the \textbf{Kleene star}, not * as Java, JavaScript, \& co. interpret * in regexps.~\cite{SO6}
% regex sample row 10
\end{quote}

Closer towards \textit{Recommendation} we place \textsc{Awareness} links that steer users' attention towards related resources, without particularly endorsing them, as well as \textsc{Reference} links that users include to make statements verifiable and more trustworthy by pointing to documentation resources supporting their claims.

One purpose of links towards the \textit{Recommendation} end of the spectrum is to explicitly guide readers to \textsc{BackgroundReading}. Such links are especially helpful for users who are new to a topic or domain since they support them in identifying relevant background knowledge:

\begin{quote}
\sf
\footnotesize
There is a good detailed description of lookarounds (look-behind and look-ahead) as well as a lot of other regex ``magic'' \textbf{here}\cite{SO23}
\end{quote}

Finally, we find explicit \textsc{Recommendation} links.
They allow readers to retrieve a specific software component recommended by a Stack Overflow author using the provided link (reproduced from Section~\ref{s:quantitative}).

\begin{quote}
\sf
\footnotesize
You could use \textbf{Apache Commons Lang} for that...~\cite{SO7}
 % There you have methods like \textbf{isNumeric} and \textbf{isAlphanumeric}~\cite{SO7}
% \emph{Android} sample row 342
\end{quote}

\par\noindent\rule{\columnwidth}{1.5pt}
\textbf{Implication:} Forum users add links to documentation for a variety of purposes. This purpose may not be clear to the reader. Links whose purpose is not clear may confuse or waste the time of inexperienced users, who are surmised to visit more links as they navigate web sites~\cite{CK2006a}. Automated analysis of link data (e.g.,~\cite{GCS2013a}) may miss opportunities for additional interpretation if link purpose is not taken into account.
\par\noindent\rule{\columnwidth}{1.5pt}
%\end{roundedbox}

\subsection*{Purpose--Resource Correspondence}

In the two cases we studied, mined association rules show consistent relations between a resource type (e.g., Wikipedia, Stack Overflow) and a link's purpose. Links to \emph{Wikipedia}, for example, often serve to define \textsc{Concept}s, an observation consistent with previous work~\cite{VJH2018a}.
Links to the documentation of software components and tools are often included to recommend the tool rather than to refer to the linked document specifically (\textsc{Recommendation}).

\par\noindent\rule{\columnwidth}{1.5pt}
\textbf{Implication:} For technology domains where certain resource types can be strongly associated with a link purpose, it may be possible to automatically recommend links to enhance a post, or infer the purpose of a linked resource.
\par\noindent\rule{\columnwidth}{1.5pt}

\subsection*{Domain-specific Link Usage}

The distribution of link purposes shown in Figure~\ref{fig:code-frequencies} and detailed in Tables~\ref{tab:codes-regex}--\ref{tab:codes-android} shows remarkable consistency between cases except for two major differences:
Purpose code \textsc{Concept} is about twice as common in case \textit{Regex} and purpose code \textsc{Awareness} is about twice as common in case \textit{Android}.
For the other codes, the relative frequency differs not more than 3 percentage points.
Both differences mentioned above are significant at the level $\alpha=0.01$ (see Section~\ref{sec:code-frequency}). From this we conjectured that the higher proportion of \textsc{Concept} links is explained by the theoretical nature of the domain, which involves concepts such as ``parsing'', ``context-free grammar'', ``pattern'', etc. This observation is corroborated by the regression model, which shows that one of the dominant features for explaining whether a Stack Overflow
question related to regular expressions will attract a particular type of
documentation link include such theoretical concepts, namely ``parsing'' and ``pattern''. As for Android, the extensive use of links for \textsc{Awareness}
purposes can be explained by the size of this technology ecosystem.

As mentioned above, we added the documentation resource \emph{Youtube} while adapting the classifier for the \textit{Android} case.
This is another manifestation of domain-specific link usage, because in the \emph{Regex} case, only 26 posts pointed to \emph{Youtube} (0.09\% of all posts containing links), while in the \emph{Android} case, linking \emph{Youtube} videos was much more common (1,822 posts or 0.8\% of all posts containing links).
The difference was significant according to a two-tailed Fisher's exact test~\cite{F1922} with a significance level of $\alpha=0.01$ ($\text{p-value}<2.2\times 10^{-16}$).
Typical use cases of linking \emph{Youtube} videos include pointing to tutorials\footnote{Example tutorial: \url{https://youtu.be/fn5OlqQuOCk}} or conference talks.\footnote{Example conference talk: \url{https://youtu.be/N6YdwzAvwOA}}

% not really documentation:
% but also providing screencasts to demonstrate how issues can be reproduced.\footnote{Exemplary screencast: \url{https://youtu.be/3mawem-refw}}

\par\noindent\rule{\columnwidth}{1.5pt}
\textbf{Implication:} Links to documentation resources are a reflection of the information needs typical to a technology domain. Details on the distribution of purpose links for a domain can thus assist in the design of documentation.
\par\noindent\rule{\columnwidth}{1.5pt}

% \begin{roundedbox}
% \textbf{Implication 3 (Utilizing domain{\scriptsize$\leftrightarrow$}purpose{\scriptsize$\leftrightarrow$}resource):}\\
% The above-mentioned correspondence between purpose and resource, together with knowledge about domain-specific link usage, can be utilized to derive and visualize the purpose of existing links and to predict the link purpose for new links.
% \end{roundedbox}

\subsection*{Missing Link Context}

Even though Stack Overflow encourages users to provide context for links~\cite{SO2019a}, they are rarely accompanied by a \textsc{Quote}~~\cite{SO2019b} or a \textsc{Summary}. Our analysis shows that, for 72.8\% of the analyzed links, authors did not provide a quote and for 83.3\% of the links they did not provide a summary. Although in some situations this lack of context may render links worthless once their target is unavailable, our analysis also revealed valid use cases for links without context, as links at the \textit{Citation} end of the purpose spectrum do not necessarily need context. However, links towards the \textit{Recommendation} end of the spectrum should always be accompanied by additional information to preserve that information in case the linked resources becomes unavailable.

\par\noindent\rule{\columnwidth}{1.5pt}
\textbf{Implication:} Our link \emph{Purpose Spectrum} observation allows us to modulate the requirement to add context for links, given that our data shows the context to be self-explanatory for links whose purpose is akin to a citation. We hypothesize that the importance of context for orienting users is proportional to a link's position on the purpose spectrum. Missing context is thus not necessarily a problem for links whose purpose is citation.
\par\noindent\rule{\columnwidth}{1.5pt}

% \begin{roundedbox}
% \textbf{Implication 2 (Context not always required):}\\
% Stack Overflow motivates why users should provide context for links, but frequently, users do not follow their advise.
% While the lack of context is not a problem for all link purposes we identified, it is often unclear for users whether the context is missing or not required.
% Again, making that information explicit would increase the quality of posts.
% \end{roundedbox}

\subsection*{Reputation-Expertise Mismatch}

The logistic regression analysis shows that users with a high reputation score are not necessarily more familiar with reference documentation than lower reputation users. Links to the \textsc{AndroidReference} documentation are associated with questions asked by users with a higher reputation. The median user reputation of users asking questions which attract links to the \textsc{AndroidReference} documentation in the dataset used for the logistic regression analysis is $1063.5$, while the corresponding median for the remaining questions is $86$. A manual inspection of the corresponding questions suggests that many of these high-reputation users are outsiders whose expertise is, based on the questions they typically answer, in areas other than Android (often iOS). Similarly, links to \textsc{Wikipedia} are also associated with questions asked by users with a higher reputation.

%\todo{R2 had difficulties understanding the following, but I have no idea how to better phrase this.}

\par\noindent\rule{\columnwidth}{1.5pt}
\textbf{Implication:}
In previous research efforts, researchers have often treated an individual's reputation on Stack Overflow as a proxy for this individual's general programming knowledge (e.g.,~\cite{MorrisonMurphyHill2013}). Our results indicate that this operationalization may not be valid in all scenarios, because Stack Overflow authors'  knowledge is domain-specific.
\par\noindent\rule{\columnwidth}{1.5pt}

% \begin{roundedbox}
% \textbf{Implication 4 (Reputation $\ne$ Domain expertise):}\\
% In the past, researchers have treated reputation on Stack Overflow as a proxy for knowledge (e.g.,~\cite{MorrisonMurphyHill2013}).
% Our results indicate that there are scenarios where this operationalization may not be valid.
% \end{roundedbox}

% \subsection*{Relationships}
%
% \todo{revise the following paragraph}
%
% Figure~\ref{fig:observations} visualizes the relationships between the findings we outlined above.
% The overarching result connecting all other findings the the purpose spectrum of Stack Overflow links.
% The domain-specific usage of links and the correspondence between documentation resources and link purposes can be seen as a consequence of this spectrum.
% The prevalent lack of link context is sometimes mitigating by a close purpose-resource correspondence, because users can identify \textbf{citation} links that are not meant to be followed based on the linked resource.
% The purpose-resource correspondence also implies that the presences of certain links in answers can be used as a proxy to identify the absence of knowledge for the user asking the corresponding question, leading to the reputation-expertise mismatch we described above.

\subsection*{Summary}

The findings described in the previous paragraphs build on each other to form a small conceptual framework defined in terms of logical implications.
Figure~\ref{fig:observations} summarizes the findings and their relationships.

Our primary finding concerns the variety of linking purposes we elicited and the observation that linking purpose types span a spectrum that characterizes to what extent a link is intended to be followed (\emph{Purpose Spectrum}).

We also collected evidence of a notable correspondence between a resource type (e.g., Wikipedia) and a link's purpose (\emph{Purpose--Resource Correspondence}), and that link usage may be specific to a technology domain (\emph{Domain-Specific Link Usage}). Both of these observations are consequences of \emph{Purpose Spectrum} in the sense that it is the observed richness of linking purposes that enables the elicitation of specific linking practices.

A fourth observation is the extent to which links in Stack Overflow threads lack context, despite the presence of guidelines explicitly requesting such context (\emph{Missing Link Context}). To a certain extent, this observed problem can be mitigated by \emph{Purpose--Resource Correspondence} because this correspondence supports partial inference of a link's purpose.

Finally, our analysis reveals a pattern that would be counter-intuitive at first glance: users with high reputation attract answers with links to the reference documentation, which can also be construed a symptom of lack of expertise (\emph{Reputation-Expertise Mismatch}). This finding is enabled by the \emph{Purpose--Resource Correspondence} which relates links to documentation resources with a type of information need. 

\section{Threats to Validity}
\label{s:threats}

\begin{figure}
\centering
\includegraphics[width=0.95\columnwidth,  trim=0.0in -0.2in 0.0in 0.0in]{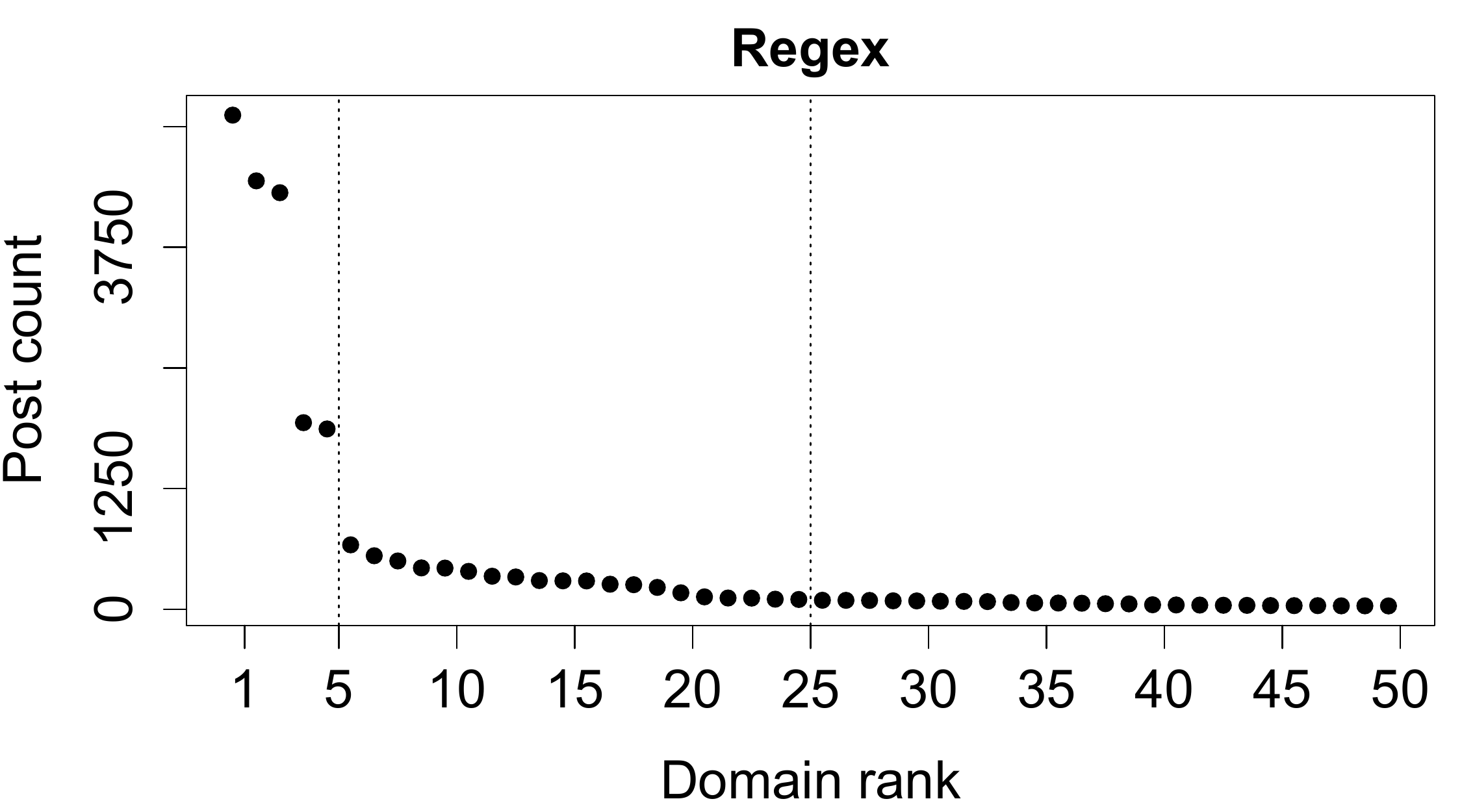} % left, bottom, right, top
\includegraphics[width=0.95\columnwidth,  trim=0.0in 0.0in 0.0in 0.0in]{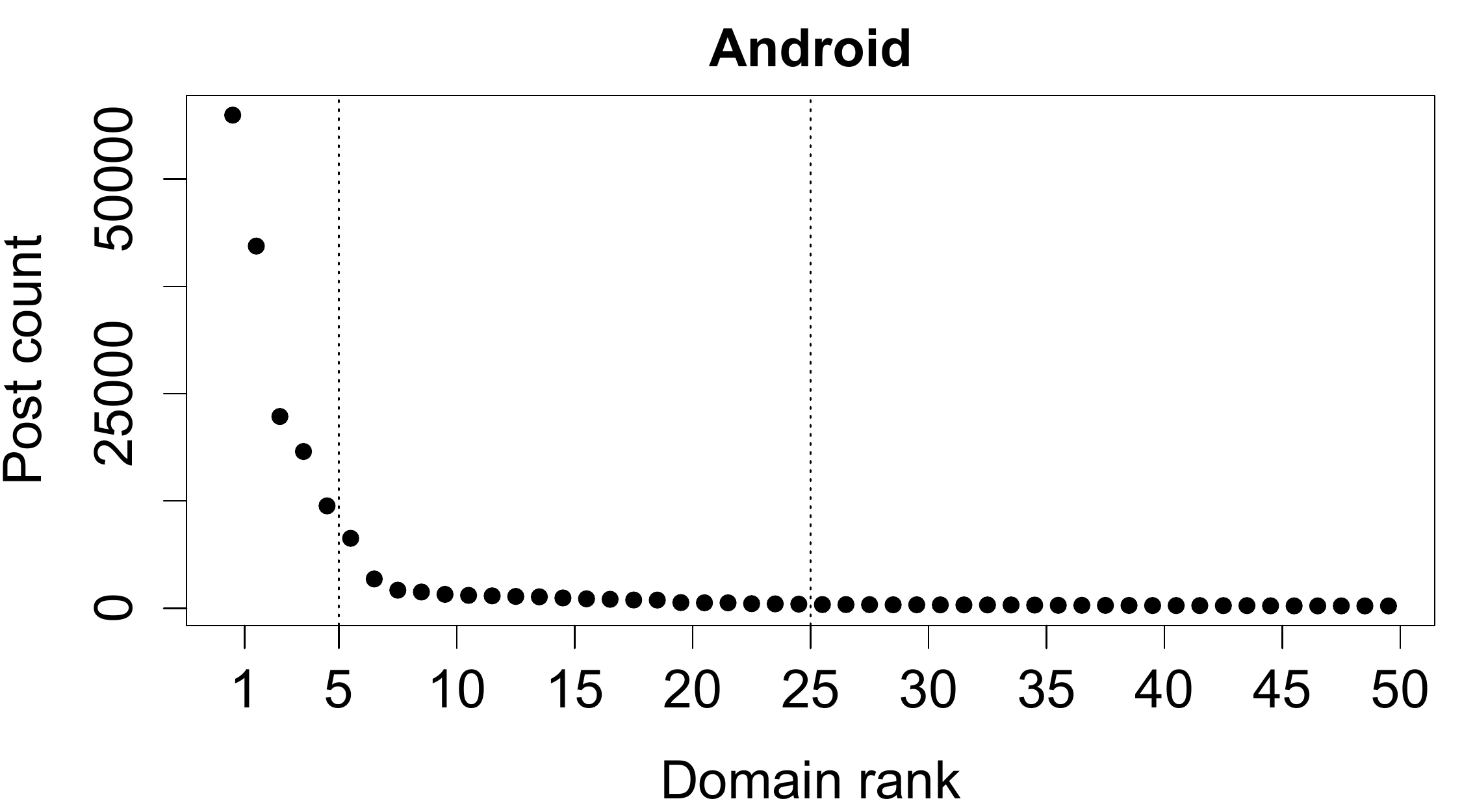} % left, bottom, right, top
\caption{Post count for top 100 root domains, top 5 shown in Tables~\ref{tab:java-regex-root-domains} and \ref{tab:java-android-root-domains}, top 25 incorporated in classifier.}
\label{fig:postcount_per_rootdomain}
\end{figure}

The external validity of our results may be limited due to our choice of the two specific domains \emph{Regex} and \emph{Android}, both of which were taken from the Java domain.
%  \todo{Do we need a reference for this?}
While Java is one of the most popular programming languages today,\footnote{TIOBE Index for December 2019, \url{https://www.tiobe.com/tiobe-index/}, verified 16 December 2019.} its documentation ecosystem may differ from other languages.
The documentation resources we identified, such as API documentation~\cite{PTG2012} and Wikipedia~\cite{VJH2018a}, are, however, likely to also play an important role for other languages and domains.

Another threat is that our URL mapper was only able to classify 78.5\% of all active links in the \emph{Regex} sample and 68.9\% of all active links in the \emph{Android} sample (see Section~\ref{s:sampling}).
Note that a classification of the remaining links would only add more documentation resources, but not invalidate the ones we have already identified.
Also, the number of posts containing a link to the corresponding root domain considerably drops after the top five (\emph{Regex}) respectively the top six (\emph{Android}) root domains (see Figure~\ref{fig:postcount_per_rootdomain}).
This also means that the marginal profit of analyzing additional root domains drops considerably after analyzing the most frequently referenced root domains.
While iteratively building the classifier, two authors continuously discussed the emerging documentation resource categories and corresponding sub-pages of the root domains.
Since we followed a whitelisting approach based on regular expressions matching certain sub-paths of the domains, making all decisions unanimously, the false positive rate of our approach is very low.
We may, however, have missed certain sub-paths, marking them as \emph{NotDocumentation} when they were in fact documentation resources (false negatives).
We mitigated this bias by manually inspecting the links marked as \emph{NotDocumentation} after each iteration, filtering out links that clearly did not point to documentation resources until no links were left to analyze.

The stratified sampling strategy we used to select documentation links for our analyses represents a threat to the external validity of our results.
Note that in a random sample, the top three documentation resources, especially the internal Stack Overflow links, would overshadow the less frequent documentation resources (see Table~\ref{tab:doc-resources}).
Our sampling strategy allowed us to analyze a broader and more diverse sample of documentation resources not dominated by those very frequent link targets.
In the association rule analysis we conducted, support and confidence only hold for our samples---they would differ in non-stratified samples.
In Section~\ref{s:quantitative}, we described how to interpret those values considering the stratification.
Moreover, the fact that all rules derived from the \emph{Regex} sample were also present in the \emph{Android} sample further supports their credibility.

The purpose distribution would likely differ in a random sample.
However, in a random sample, frequently referenced documentation resources such as \emph{Stack Overflow}, \emph{JavaAPI}, and \emph{AndroidAPI} would dominate the analysis.
The stratification allowed us to consider a more diverse range of resources and purposes.

Qualitative data analysis always depends on the imagination and perception of the researcher.
To mitigate this threat, all three authors conducted the qualitative analysis independently.
We coded links in sets of up to 100 links and thoroughly discussed our results after finishing each set.
After assessing our inter-rater agreement ($\kappa$ values between 0.65 and 0.80, see Table~\ref{tab:cohen}), we only assigned a code if at least two researchers agreed on it.

% (see description of coding process in Section~\ref{s:sampling}).

%Finally, we cannot claim that a replication of our study in the future would identify the exact same results as new ways of documentation emerge and others are abandoned.

% \input{s9-relwork}
\section{Conclusion}
\label{s:conclusion}

%\todo{Finding and conclusions: We will use R2's feedback to improve the structure, presentation, and content of our findings and the corresponding conclusions. Considering R3's feedback, we will also revise those sections to better reflect practical implications.}

%\todo{Highlight practical implications, connect results again to improving the efficiency of information flow, see comment in introduction.}

Over the past decade, the community question answering platform Stack Overflow has become extremely popular among programmers for finding and sharing knowledge. However, the site does not exist in isolation, and users frequently link to other documentation sources, such as API documentation and encyclopedia articles, from within questions, answers, or comments on Stack Overflow. To understand how and why documentation is referenced from Stack Overflow threads, we conducted a multi-case study of links in two different technology domains, regular expressions and Android development. We used qualitative and quantitative research methods to systematically investigate the context and purpose of a sample of 759 documentation links.

We identified a spectrum of purposes for which links are included in Stack Overflow threads, ranging from \textsc{Attribution} and \textsc{LinkedMention} on the citation end of the spectrum to \textsc{BackgroundReading} and \textsc{Recommendation} of software artifacts on the recommendation side.
% (see Figure~\ref{fig:purpose-spectrum}).
Citations are not necessarily meant to be consulted whereas recommendations are explicit requests to follow a link.
This observation relates to Stack Overflow's recommendation to add context to every link: While adding context in the form of summaries or quotes is important for links on the recommendation end of the purpose spectrum, it is less important for links primarily included for citation purposes.

We also found that links to documentation resources are a reflection of the information needs typical to a technology domain. For example, \textsc{Concept} links were twice as common in threads about regular expressions compared to Android, while we found the opposite for \textsc{Awareness} links. These insights can inform the design and customization of documentation for different technology domains.

Our work forms a first step towards understanding how and why documentation resources are referenced on Stack Overflow, with the ultimate goal of improving the efficiency of information diffusion between Stack Overflow and the broader software documentation ecosystem, as motivated in Section~\ref{s:examples}. In the short term, Stack Overflow authors can use our results to reflect on the intended purpose before posting a link, and to learn how they can make their post more valuable by providing context.
% Our future work includes distilling our findings into recommendations for Stack Overflow authors and to develop techniques that classify the context and purpose of a link in a Stack Overflow thread to automatically issue recommendations about how critical it is to follow a particular link.

%\todo{Future work: refinement of our coding guide by adding sub-codes for the purposes}
%\todo{Future work: detect recommendations in thread, aggregate and visualize them for the whole SO thread?}

Another direction for future work is developing tool support for guiding Stack Overflow users to enhance (potential) information diffusion.
One tool could assist readers of Stack Overflow threads by automatically classifying links in posts or comments along the purpose spectrum we presented in this paper.
Such a tool could be implemented as a browser plugin visualizing the determined purpose of the link, helping users to judge whether the link it is worth following based on their particular needs.
Another idea is to extend the models we presented in Section~\ref{s:quantitative} to be able to recommend Stack Overflow authors to include a certain type of link while creating or revising Stack Overflow posts.

%\appendices

%\ifCLASSOPTIONcompsoc
  %\section*{Acknowledgments}
%\else
  %\section*{Acknowledgment}
%\fi

\ifCLASSOPTIONcaptionsoff
  \newpage
\fi

\bibliographystyle{IEEEtran}
\bibliography{literature}


\begin{IEEEbiography}[{\includegraphics[width=1in,height=1.25in,clip,keepaspectratio]{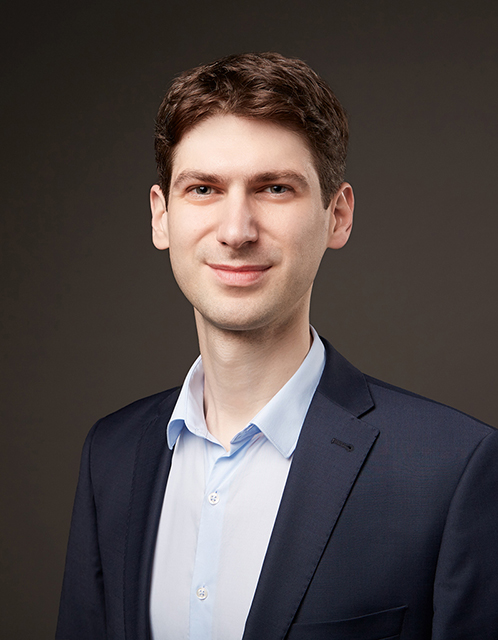}}]{Sebastian Baltes}
is a Lecturer in the School of Computer Science at the University of Adelaide, Australia.
He received his Ph.D. degree in Computer Science from the University of Trier, Germany.
In his research, he empirically analyzes software developers' work habits to derive tool requirements and to identify potential process improvements.
Most empirical studies he conducts combine qualitative and quantitative research methods; he is especially interested in interdisciplinary research, including the application of theories and methods from other disciplines in the software engineering domain.
\end{IEEEbiography}

\begin{IEEEbiography}[{\includegraphics[width=1in,height=1.25in,clip,keepaspectratio]{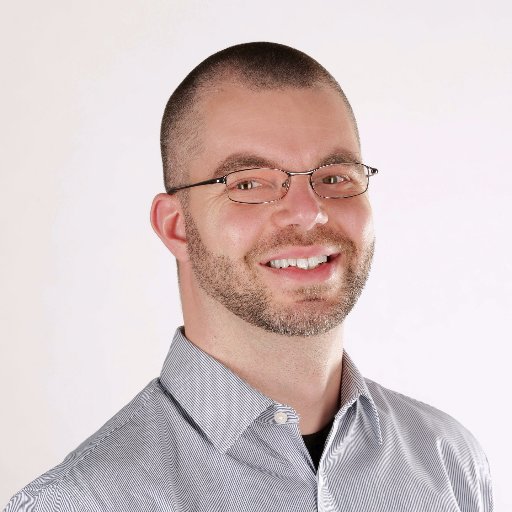}}]{Christoph Treude}
is an ARC DECRA Fellow and a Senior Lecturer in the School of Computer Science at the University of Adelaide, Australia.
He received his Ph.D. in computer science from the University of Victoria, Canada in 2012.
The goal of his research is to advance collaborative software engineering through empirical studies and the innovation of tools and processes that explicitly take the wide variety of artifacts available in a software repository into account.
He currently serves on the editorial board of the Empirical Software Engineering journal and as general co-chair for the IEEE International Conference on Software
Maintenance and Evolution 2020.
\end{IEEEbiography}

\begin{IEEEbiography}[{\includegraphics[width=1in,height=1.25in,clip,keepaspectratio]{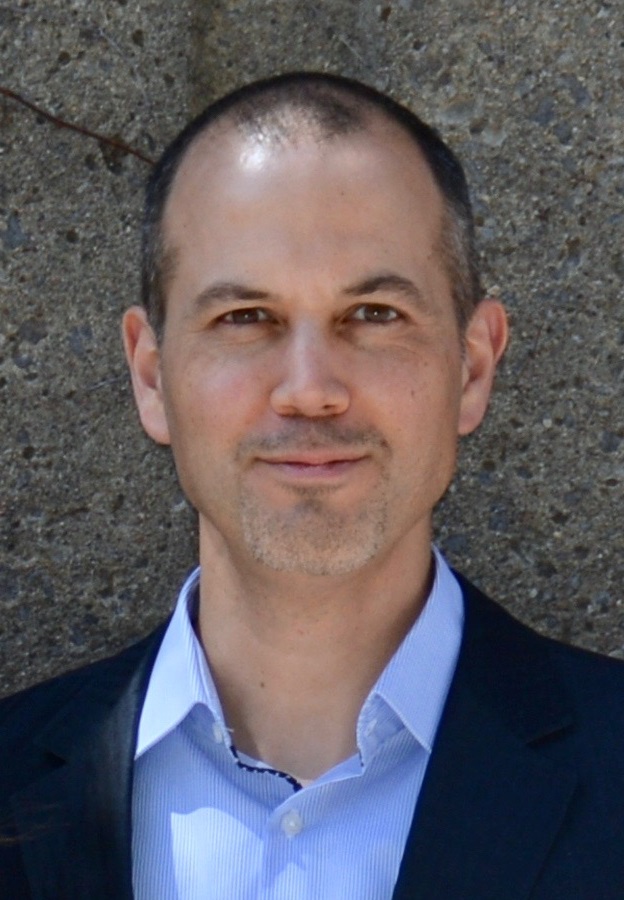}}]{Martin Robillard}
is a Professor of Computer Science at McGill University.
His research investigate how to facilitate the discovery and acquisition of technical, design, and domain knowledge to support the development of software systems.
He served as the Program Co-Chair for the 20th ACM SIGSOFT International Symposium on the Foundations of Software Engineering (FSE 2012) and the 39th ACM/IEEE International Conference on Software Engineering (ICSE 2017).
He received his Ph.D. and M.Sc. in Computer Science from the University of British Columbia and a B.Eng. from \'{E}cole Polytechnique de Montr\'{e}al.
\end{IEEEbiography}



\vfill


\end{document}